\documentclass[amsmath,amssymb,
 aps,prd,reprint,onecolumn,showpacs,showkeys]{revtex4-2}
\pdfoutput=1
\usepackage{graphicx}
\usepackage{dcolumn}
\usepackage{bm}
\usepackage{hyperref}
\usepackage{array}
\usepackage{float}
\usepackage{textcomp}
\usepackage{mathtools}
\usepackage{multirow}
\usepackage{amsfonts}
\usepackage{bm}
\usepackage{wasysym}
\PassOptionsToPackage{normalem}{ulem}
\usepackage{ulem}

\begin{document}

\title{Growth of Matter Perturbations in the Bi-Galileons Field Model}%
\author{Khireddine Nouicer }%
\email{khnouicer@univ-jijel.dz}
\author{Hamza Boumaza }%
\email{boumaza14@yahoo.com}
\affiliation{Laboratory of Theoretical Physics and Department of Physics,\\
Faculty of Exact and Computer Sciences,\\ University Mohamed Seddik Ben Yahia,\\
BP 98, Ouled Aissa, Jijel 18000, Algeria}
\date{December 2018}%


\begin{abstract}
We study a dark energy cubic bi-Galileons field model based
on truncation of the recently proposed generalized covariant multi-Galileons model. We investigate the cosmological
 dynamic of the model by the theory of dynamical systems through
 the analysis of the properties of the fixed points in each cosmological epoch.
 We show the existence of two tracker solutions, one of which is that of the cubic single Galileon model and the other solution is
the signature of the second Galileon field. Exploiting the competition between the two Galileon fields,
 we find a dark energy solution that avoids the approach to the tracker solution with dark energy equation of state $w_{DE}=-2$
during the matter epoch which is disfavored by the observational data. We study also the growth rate of matter perturbations.
 Using recent  $f\sigma_{8}$ redshift space distortion (RSD) and model-independent observational Hubble (OHD) data sets,
we place observational constraints on the coupling constant
and cosmological parameters of the bi-Galileons model through Monte Carlo numerical method based on the Metropolis-Hastings algorithm.
 We find that the amplitude of growth matter fluctuations is consistent with the Planck15 data and ease the tension between early and later clustering, and  fits better the data from the DES survey over the data from KiDS-450 survey. We also find that the best fit value for the Hubble constant is compatible with new measurements of Cepheid-supernovae distance scale.
 Finally, we perform a model selection through the Bayes factor and found that the bi-Galileons model is disfavored in
comparison to the $\Lambda$CDM model, but slightly preferred to $w$CDM model.
\end{abstract}



\maketitle

\section{Introduction}

The origin of the late time accelerated expansion of the universe,
discovered two decades ago \cite{Riess1,*Riess2,*Perlmutter}, remains
one of the theoretical challenging enigmas of modern cosmology. The
observational data \cite{Spergel,ade,aghanim2018}, obtained with
high precision, of different cosmological parameters of the standard
cosmological model, the $\Lambda$CDM model, have confirmed this fact,
and that the accelerated expansion is driven by a dark energy (DE)
component with an equation of state $w_{DE}$ close to $-1.$ Even
though the $\Lambda$CDM model has been successful in explaining the
dynamics of the universe at large scales ( particularly at the background
level), it suffers from conceptual problems like \textit{fine tuning,}
\textit{coincidence problem and the origin of dark energy }\cite{coincidence-problem}\textit{.
}In fact, the $\Lambda$CDM model describes the dark energy component
as a cosmological constant $\Lambda$, attributed to the vacuum
energy density, but its value is extremely small compared to quantum
field theory calculation \cite{quantum-correction}. Recently, two
problems came to revive the debates about the theoretical foundations
of the the $\Lambda$CDM model. The first one is the persistent tension
between the values of Hubble constant $H_{0}$ constrained from the
Cosmic Microwave Background (CMB) physics \cite{ade,aghanim2018}
and local measurements from supernovae distance scale \cite{WLFreedman, BVRiess,  AdRiess}
and lensing time delays \cite{H0LiCOW}. The second more debated problem
of the $\Lambda$CDM model is the tension between the large $f\sigma_{8}$
values predicted by Planck/$\Lambda$CDM data, indicating a high level
of structure clustering, and the smaller values from the redshift
space distorsion (RSD) data obtained from galaxy redshift surveys
in the late universe.

In order to solve these problems and particularly to account for the
present accelerated expansion and the origin of dark energy driven
it, alternative models have been introduced modeling the dark energy
using scalar fields like: quintessence \cite{quintessence1,quintessence2,quintessence3,quintessence4},
k-essence \cite{k-essence0,k-essence} and Brans-Dicke theories \cite{branse-dicke0,branse-dicke,branse-dicke2},
Covariant Galileon \cite{cov-gal1,cov-gal2} and Kinetic Gravity Braiding
\cite{kgb1,kgb2}. These models, which are large scale modification
of general relativity, are actually known as sub-classes of the most
general scalar-tensor theory with second order equations of motion,
derived by Horndeski \cite{horndeski}. The current status of Horndeski\textquoteright s
theory and beyond is reviewed in \cite{Horndesky-review}. In Horndeski
theory, a rich variety of dark energy behaviors allow deviation from
$w_{DE}=-1$ at the background level. As an example, the covariant
Galileon model and its extension extensively studied and constrained
in Ref.\cite{obesr-cov-gal,cov-gal-3,observation-extended-cov-gal-1,observation-extended-cov-gal-2,cov-gal-41,cov-gal-4,cov-gal-5,cov-gal-6,Tsujikawa-Felice},
showed the existence of a solution, known as tracker solution, that
approach the Sitter solution at the late time universe. Despite its
simplicity, this tracking solution is ruled out from the joint data
analysis using Supernovae Ia (SNIa), Cosmic Microwave Background (CMB) and Baryon Acoustic Oscillation (BAO)
\cite{obesr-cov-gal} due to the bad behavior of the DE equation of
state $w_{DE}=-2$ in the matter epoch. In addition, the insertion
of nonlinear terms in the cubic Galileon action, such as the condensate
Galileon model (CGM), modifies the evolution of DE equation of state
and lead to a model favored over $\Lambda$CDM model by the criterion
of Bayesian model selection \cite{Galileon-condensate,condensate-gal}.

A recent alternative to scalar-tensor theory is the bi-scalar-tensor
theory in which the action contains two scalar fields rather than
one. These kinds of modified gravity models have been studied in the
flat space-time at the background level \cite{bi-gal-first,cs} and
perturbed space-time \cite{david}. Later, the generalized multi-Galileons
was proposed as multi-fields generalization of Horndeski scalar-tensor
theory following Horndeski\textquoteright s recipe, where all possible
terms appearing in the second-order field equations of the bi-scalar-tensor
theory was determined \cite{genalized-Galileon}.

In the present paper, inspired by these approaches, we will present
a sub-class of the multi-Galileons model, the bi-Galileons (BG) model,
and investigate its cosmological evolution at the background and linear
perturbed levels. We will show that the BG model can realize a variety
of dark energy equation of state depending on the initial conditions
of dynamical variables of the model. Particularly, we prospect the
viability of the BG model in the light of the recent rate growth of
matter perturbation data and Hubble parameter measurements and detect
the signature of the second field in the cosmological behavior of
the model.

The present paper is structured as follow. In Sec. \ref{sec:The-model},
we present the covariant BG model up to cubic term with constant coupling
functions, and derive the main background field equations. In Sec.
\ref{sec:Dynamical-analysis}, we study a simplified version of the
BG model using the dynamical system approach through the introduction
of suitable dimensionless variables. We perform a detailed analysis
of the stability of the fixed points in each cosmological epoch, and
investigate the existence of late-time attractor solutions. In Sec.
\ref{sec:Analysis-of-the-fixed-points} we analyze the different cosmological
implications of the fixed points on the behavior of the dark equation
of state. Particularly, we focus on a dark energy solution with a
dark energy equation of state close the $\Lambda$CDM model. In Sec.
\ref{sec:Growth-rate-of} we study the evolution of the growth rate
of matter perturbations in the quasi-static approximation on sub-horizon
scales, and compute the equations governing the evolution of perturbations.
In Sec. \ref{sec:Observational-constraints} we perform a parameter
estimate using Monte Carlo analyses and confront our expectations
with the $\Lambda$CDM and $w$CDM models. Finally, Sec.\ref{sec:Conclusion}
is devoted to conclusions.

\section{The Bi-Galileons model\label{sec:The-model}}

We consider the following scalar-tensor modified model of gravity
with two Galileons fields $\varphi^{I}\left(I=1,\,2\right)$, labeled
BG model, in a four-dimensional spacetime
\begin{eqnarray}
S & = & \int dx^{4}\sqrt{-g}\Bigl(\frac{R}{2}+a_{IJ}X^{IJ}-b_{KIJ}X^{IJ}\varphi_{;\mu}^{K;\mu}+\mathcal{L}_{m}+\mathcal{L}_{r}\Bigr),\label{S-1}
\end{eqnarray}
where $R$ is Ricci scalar and the dimensionless constants $a_{IJ}\,\textrm{and}\,b_{KIJ}\left(I,\,J,\,K=1,\,2\right)$
are symmetric in $I$, $J$ . The Lagrangian $\mathcal{L}_{m}$ and
$\mathcal{L}_{r}$ stands as usual for  matter and radiation fields,
respectively. The notation $X_{IJ}\equiv-\frac{1}{2}g^{\mu\nu}\partial_{\mu}\varphi^{I}\partial_{\nu}\varphi^{J}$
is the kinetic term for $\varphi^{I}$ when $I=J$ and represents
the coupling between the field velocities of the fields when $I\neq J$
. The action (\ref{S-1}) is a particular truncation of the more general
covariant extension of the Galileon field model addressed in \cite{GG-inflation},
and is invariant under the shift transformation $\varphi^{I}\rightarrow\varphi^{I}+const$..
The usual action for the cubic single Galileon field, labeled SG
model, is recovered by setting $a_{2J}=0$ and $b_{2JK}=0$.

Varying the action with respect to the metric gives the field equations:
\begin{eqnarray}
G_{\mu\nu}-a_{IJ}\varphi_{;\mu}^{I}\varphi_{;\nu}^{J}-b_{KIJ}\left(\varphi^{I;\alpha}\varphi_{;\mu}^{J}\varphi_{;\alpha\nu}^{K}+\varphi^{I;\alpha}\varphi_{;\nu}^{J}\varphi_{;\alpha\mu}^{K}-\varphi_{;\mu}^{I}\varphi_{;\nu}^{J}\varphi_{;\mu}^{K;\mu}\right)\nonumber \\
-g_{\mu\nu}\left(a_{IJ}X^{IJ}-b_{IJK}\varphi^{I;\alpha}\varphi^{J;\beta}\varphi_{;\alpha\beta}^{K}\right) & = & T_{\mu\nu}^{(m)}+T_{\mu\nu}^{(r)}\quad\quad\label{EE}
\end{eqnarray}
where $G_{\mu\nu}$ is the Einstein tensor and $T_{\mu\nu}^{(m)}$
and $T_{\mu\nu}^{(r)}$ are the matter and radiation energy-moment
tensors, respectively. For a perfect fluid we have
\begin{eqnarray}
T_{\mu\nu}^{(i)} & = & (\rho_{i}+P_{i})u_{\mu}u_{\nu}+g_{\mu\nu}P_{i},\label{energy-tensor-moment}
\end{eqnarray}
where $u_{\mu}$, $\rho_{i}$ and $P_{i}$ are the
four velocity vector, energy density and pressure of the fluid, respectively. The
energy-momentum conservation law is provided by
\begin{equation}
\nabla_{\mu}T^{(i)\mu\nu}=0.\label{eq:contunu}
\end{equation}
Now, varying the action (\ref{S-1}) with respect to the scalar field
$\varphi^{I}$ leads to the appearance of third order derivative in
the fields, which are canceled by imposing the constraint $b_{IJK}=b_{JKI}.$
Then we are let with second order field equation
\begin{equation}
J_{I;\mu}^{\mu}=0\label{eq:BG_f1}
\end{equation}
where
\begin{eqnarray}
J_{I}^{\mu}=a_{IJ}\varphi^{J;\mu}-b_{IJK}\left(\varphi^{K;\mu}\Square\varphi^{J}+X^{JK;\mu}\right) & .\label{eq:BG_f2}
\end{eqnarray}
We note the presence of first order derivative of $\varphi^{I}$ which
breaks the Galilean symmetry $\partial_{\mu}\varphi^{I}\rightarrow\partial_{\mu}\varphi^{I}+b_{\mu},\;\varphi^{I}\rightarrow\varphi^{I}+c$.

Let us now study background cosmological solutions of the BG model.
We consider the spatially flat Friedmann-Robertson-Walker (FRW) metric

\begin{equation}
ds^{2}=-dt^{2}+a\left(t\right)^{2}\delta_{ij}dx^{j}dx^{i},\label{ds^2}
\end{equation}
where $a\left(t\right)$ is the scale factor. Substituting this ansatz
in Einstein equations (\ref{EE}) we obtain the Friedmann equations
\begin{eqnarray}
3H^{2} & = & a_{IJ}X^{IJ}+6b_{KIJ}HX^{IJ}\dot{\varphi}^{K}+\rho_{m}+\rho_{r},\label{FRW-1}\\
-\left(3H^{2}+2\dot{H}\right) & = & a_{IJ}X^{IJ}-2b_{KIJ}X^{IJ}\ddot{\varphi}^{K}+p_{m}+p_{r},\label{FRW-2}
\end{eqnarray}
where $H=\dot{a}/a$ is the Hubble parameter. From Eqs.(\ref{FRW-1}) and (\ref{FRW-2}),
 we identify the dark energy density
and pressure as
\begin{eqnarray}
\rho_{DE} & = & a_{IJ}X^{IJ}+6b_{KIJ}HX^{IJ}\dot{\varphi}^{K},\label{ro-phi}\\
P_{DE} & = & a_{IJ}X^{IJ}-2b_{KIJ}X^{IJ}\ddot{\varphi}^{K}.\label{p-phi}
\end{eqnarray}
We also define the dark energy equation of state by
\begin{equation}
\omega_{DE}=\frac{P_{DE}}{\rho_{DE}}=\frac{a_{IJ}X^{IJ}-2b_{KIJ}X^{IJ}\ddot{\varphi}^{K}}{a_{IJ}X^{IJ}+6b_{KIJ}HX^{IJ}\dot{\varphi}^{K}}.\label{EoS-field}
\end{equation}
It is clear that $w_{DE}$ can cross the phantom divide line, $w_{DE}=-1.$

Inserting the metric (\ref{ds^2}) into (\ref{eq:BG_f1}), we obtain
the scalar fields equations of motion
\begin{equation}
a_{IJ}\left(\ddot{\varphi}^{J}+3H\dot{\varphi}^{J}\right)+3b_{IJK}\left(\left(3H^{2}+\dot{H}\right)\dot{\varphi}^{J}\dot{\varphi}^{K}+H\dot{\varphi}^{(K}\ddot{\varphi}^{J)}\right)=0,\label{FRW-scalarfield}
\end{equation}
where $\dot{\varphi}^{(K}\ddot{\varphi}^{J)}$ stands for the corresponding
symmetrized quantity.

\section{Dynamical analysis and cosmological evolution\label{sec:Dynamical-analysis}}

In this section we are interested by the homogeneous and isotropic
cosmology of a simple version of the the model introduced in the last
section. We consider a BG model where among the coupling constants
in (\ref{S-1}) we choose
\begin{equation}
a_{12}=a_{21}=b_{222}=0.\label{constant}
\end{equation}
For this model the Friedmann equations simplify to
\begin{eqnarray}
3H^{2} & = & \frac{a_{11}(\dot{\varphi}^{1}){}^{2}}{2}+\frac{a_{22}(\dot{\varphi}^{2}){}^{2}}{2}+3H\left(b_{111}(\dot{\varphi}^{1}){}^{3}+3b_{211}\dot{\varphi}^{2}(\dot{\varphi}^{1}){}^{2}+3b_{122}(\dot{\varphi}^{2}){}^{2}\dot{\varphi}^{1}\right)\nonumber \\
 &  & +\rho_{m}+\rho_{r},\label{frw1}\\
-\left(3H^{2}+2\dot{H}\right) & = & \frac{a_{11}(\dot{\varphi}^{1}){}^{2}}{2}+\frac{a_{22}(\dot{\varphi}^{2}){}^{2}}{2}-b_{111}(\dot{\varphi}^{1}){}^{2}\ddot{\varphi}^{1}-b_{211}\left(2\dot{\varphi}^{2}\dot{\varphi}^{1}\ddot{\varphi}^{1}+(\dot{\varphi}^{1}){}^{2}\ddot{\varphi}^{2}\right)\nonumber \\
 &  & -b_{122}\left(2\dot{\varphi}^{2}\dot{\varphi}^{1}\ddot{\varphi}^{2}+(\dot{\varphi}^{2}){}^{2}\ddot{\varphi}^{1}\right)+\frac{\rho_{r}}{3}.\label{frw2-1}
\end{eqnarray}
where the density and pressure of dark energy are given by
\begin{flalign}
\rho_{DE}= & \frac{a_{11}(\dot{\varphi}^{1}){}^{2}}{2}+\frac{a_{22}(\dot{\varphi}^{2}){}^{2}}{2}+3H\left(b_{111}(\dot{\varphi}^{1}){}^{3}+3b_{211}\dot{\varphi}^{2}(\dot{\varphi}^{1}){}^{2}+3b_{122}(\dot{\varphi}^{2}){}^{2}\dot{\varphi}^{1}\right),\quad\\
P_{DE}= & \frac{a_{11}(\dot{\varphi}^{1}){}^{2}}{2}+\frac{a_{22}(\dot{\varphi}^{2}){}^{2}}{2}-b_{111}(\dot{\varphi}^{1}){}^{2}\ddot{\varphi}^{1}-b_{211}\left(2\dot{\varphi}^{2}\dot{\varphi}^{1}\ddot{\varphi}^{1}+(\dot{\varphi}^{1}){}^{2}\ddot{\varphi}^{2}\right)\nonumber \\
 & -b_{122}\left(2\dot{\varphi}^{2}\dot{\varphi}^{1}\ddot{\varphi}^{2}+(\dot{\varphi}^{2}){}^{2}\ddot{\varphi}^{1}\right),
\end{flalign}
and verify the continuity equation
\begin{equation}
\dot{\rho}_{DE}+3H\left(\rho_{DE}+P_{DE}\right)=0.
\end{equation}
The BG fields equations of motion on the FRW background reads as
\begin{align}
a_{11}\left(\ddot{\varphi}^{1}+3H\dot{\varphi}^{1}\right)+3b_{111}\left(\left(3H^{2}+\dot{H}\right)(\dot{\varphi}^{1}){}^{2}+2H\ddot{\varphi}^{1}\dot{\varphi}^{1}\right)+3b_{122}\biggl(\left(3H^{2}+\dot{H}\right)(\dot{\varphi}^{2}){}^{2}\nonumber \\
+2H\ddot{\varphi}^{2}\dot{\varphi}^{2}\biggr)+6b_{211}\left(\left(3H^{2}+\dot{H}\right)\dot{\varphi}^{1}\dot{\varphi}^{2}+H\left(\dot{\varphi}^{2}\ddot{\varphi}^{1}+\dot{\varphi}^{1}\ddot{\varphi}^{2}\right)\right) & =0,\quad\label{eqfilde1}\\
a_{22}\left(\ddot{\varphi}^{2}+3H\dot{\varphi}^{2}\right)+3b_{211}\left(\left(3H^{2}+\dot{H}\right)(\dot{\varphi}^{1}){}^{2}+2H\ddot{\varphi}^{1}\dot{\varphi}^{1}\right)+6b_{122}\Bigl(\left(3H^{2}+\dot{H}\right)\dot{\varphi}^{1}\dot{\varphi}^{2}\nonumber \\
+H\left(\ddot{\varphi}^{1}\dot{\varphi}^{2}+\dot{\varphi}^{1}\ddot{\varphi}^{2}\right)\Bigr) & =0.\label{eqfilde2}
\end{align}
The coupling between the two Galileons is controlled
by the coefficients $b_{111},\,b_{122}$ and $b_{211}.$ In order
to reduce the dimension of the parameter space we assume the existence
of de Sitter (dS) epoch where $H=H_{ds}$, $\dot{\varphi}^{1}=\dot{\varphi}_{ds}^{1}=u_{ds}$,
and $\dot{\varphi}^{2}=\dot{\varphi}_{ds}^{2}=v_{ds}$, where $H_{dS},\,u_{dS}$
and $v_{dS}$ are constants that can be fixed by the phase space properties
of the model. During dS epoch, Eqs. (\ref{frw2-1}), (\ref{eqfilde1})
and (\ref{eqfilde2}) are easily solved and lead to
\begin{equation}
a_{11}=3\,\frac{H_{{\it ds}}\left(2\,u_{{\it ds}}v_{{\it ds}}^{2}b_{122}-u_{{\it ds}}^{2}v_{{\it ds}}b_{211}-2\,H_{{\it ds}}\right)}{u_{{\it ds}}^{2}},\:a_{22}=-3\,\frac{H_{{\it ds}}u_{{\it ds}}\left(2\,b_{122}v_{{\it ds}}+b_{{211}}u_{{\it ds}}\right)}{v_{{\it ds}}},\quad
\end{equation}
and
\begin{flushleft}
\begin{equation}
b_{111}=-\frac{3\,u_{{\it ds}}v_{{\it ds}}^{2}b_{122}+3\,u_{{\it ds}}^{2}v_{{\it ds}}b_{211}-2\,H_{{\it ds}}}{u_{{\it ds}}^{3}}.
\end{equation}
A further reduction of the space of parameters is carried by imposing
$a_{11}=0$ and then solve for $b_{122}$ to get
\begin{equation}
b_{111}=-\frac{1}{2}\,\frac{H_{{\it ds}}\left(2+3\,\alpha\right)}{u_{{\it ds}}^{3}},\quad a_{22}=-6\,\frac{H_{{\it ds}}^{2}}{v_{{\it ds}}^{2}}
\end{equation}
where $\alpha$ is defined by $b_{211}=\frac{\alpha H_{ds}}{u_{ds}^{2}v_{ds}}.$
Further simplifications are obtained by the redefinition of the coupling
parameters as
\begin{flalign}
b_{211}\rightarrow b_{211}\frac{u_{ds}^{2}v_{ds}}{H_{ds}}=\alpha,\:b_{122}\rightarrow b_{122}\frac{u_{ds} v_{ds}^{2}}{H_{ds}}=1-\frac{\alpha}{2},\\
b_{111}\rightarrow b_{111}\frac{u_{ds}^{3}}{H_{ds}}=-1-\frac{3}{2}\alpha,\:a_{22}\rightarrow a_{22}\frac{v_{ds}^{2}}{H_{ds}^{2}}=-6.
\end{flalign}
This redefinition does not affect the dynamics and allows us to hide
the arbitrary parameters $H_{ds},\:u_{ds}$ and $v_{ds}$.
\par\end{flushleft}

\begin{flushleft}
We now introduce the dimensionless variables
\begin{equation}
r_{1}=\left(H\dot{\varphi}^{1}\right)^{-1},\,r_{2}=H^{-1}\left(\dot{\varphi}^{1}\right)^{3},\,r_{3}=\frac{\dot{\varphi}^{2}}{\dot{\varphi}^{1}},\,\Omega_{m}=\frac{\rho_{m}}{3H^{2}},\,\Omega_{r}=\frac{\rho_{r}}{3H^{2}},\label{variable}
\end{equation}
along with the notations
\begin{equation}
\epsilon_{H}=\frac{\dot{H}}{H^{2}},\,\epsilon_{\varphi_{I}}=\frac{\ddot{\varphi}^{I}}{H\dot{\varphi}^{I}}.\label{variabl}
\end{equation}
Solving Eqs.(\ref{variable}) we obtain
\begin{equation}
H=r_{1}^{-3/4}r_{2}^{-1/4},\quad\dot{\varphi}^{1}=\left(\frac{r_{2}}{r_{1}}\right)^{1/4},\quad\dot{\varphi}^{2}=r_{3}\left(\frac{r_{2}}{r_{1}}\right)^{1/4}.\label{eq:h_phi1-2}
\end{equation}
In terms of these variables, the Friedmann and the scalar field equations
read as
\begin{flalign}
4\epsilon_{H}+\left(r_{3}^{2}(\alpha-2)-4r_{3}\alpha+3\alpha+2\right)r_{2}\epsilon_{\varphi_{1}}+\left(2r_{3}\left(\alpha-2\right)-2\alpha\right)r_{3}r_{2}\epsilon_{\varphi_{2}}-6r_{1}r_{2}r_{3}^{2} & \nonumber \\
+2\bigl(\Omega_{r}+3\bigr) & =0\label{eq:Fr2}\\
\left(3\alpha+2-4r_{3}\alpha+r_{3}^{2}(\alpha-2)\right)\epsilon_{H}+2\left(2-(2r_{3}-3)\alpha\right)\epsilon_{\varphi_{1}}-2\Bigl((2-\alpha)r_{3}^{2}+2r_{3}\alpha\Bigr)\epsilon_{\varphi_{2}}\nonumber \\
+3r_{3}^{2}(\alpha-2)-12r_{3}\alpha+9\alpha+6 & =0\label{eq:sclarfield1}\\
\left((6-3\alpha)r_{3}+3\alpha\right)\epsilon_{H}+\left((6-3\alpha)r_{3}+6\alpha\right)\epsilon_{\varphi_{1}}+\left(6-3\alpha-6r_{1}\right)r_{3}\epsilon_{\varphi_{2}}-9r_{3}(\alpha-2)\nonumber \\
-18r_{3}r_{1}+9\alpha & =0\label{eq:scalarfield2}
\end{flalign}
 For completeness we solve these equations in terms of $\epsilon_{H},\:\epsilon_{\varphi_{1}}$
and $\epsilon_{\varphi_{2}}$
\begin{flalign}
\epsilon_{H} & =\frac{1}{\mathcal{D}}\left[\left(-9\,\alpha^{3}+54\,\alpha^{2}-108\,\alpha+\left(-18\,{\alpha}^{2}+72\,\alpha-72\right)r_{{1}}+72\right){r_{{3}}}^{4}\right.\nonumber \\
 & +\left(36\,{\alpha}^{3}-48\,\alpha\,{r_{{1}}}^{2}-144\,{\alpha}^{2}+144\,\alpha\right)r_{3}^{3}\nonumber \\
 & +\left(-54\,{\alpha}^{3}+84\,{\alpha}^{2}+24\,\alpha+\left(72\,\alpha+48\right)r_{{1}}^{2}+\left(108\,{\alpha}^{2}-144\,\alpha-144\right)r_{{1}}+48\right)r_{{3}}^{2}\nonumber \\
 & +\left(36\,{\alpha}^{3}+48\,{\alpha}^{2}+48\,\alpha+\left(-144\alpha^{2}-96\,\alpha\right)r_{{1}}\right)r_{{3}}+\left(54\,{\alpha}^{2}+72\,\alpha+24\right)r_{{1}}\nonumber \\
 & \left.-9\,{\alpha}^{3}-42\,{\alpha}^{2}-60\,\alpha-24\right]r_{{2}}+\left(4\,{\alpha}^{2}\Omega_{{r}}+12\,{\alpha}^{2}-16\,\alpha\,\Omega_{{r}}-48\,\alpha+16\,\Omega_{{r}}+48\right)r_{{3}}^{2}\nonumber \\
 & +\left(-24\,{\alpha}^{2}-8\,{\alpha}^{2}\Omega_{{r}}+16\,\alpha\,\Omega_{{r}}+\left(16\,\alpha\,\Omega_{{r}}+48\,\alpha\right)r_{{1}}+48\,\alpha\right)r_{{3}}+48+12\,{\alpha}^{2}+4\,{\alpha}^{2}\Omega_{{r}}+16\,\alpha\,\Omega_{{r}}\quad\nonumber \\
 & +\left.\left(-24\,\alpha\,\Omega_{{r}}-72\,\alpha-16\,\Omega_{{r}}-48\right)r_{{1}}+48\,\alpha+16\,\Omega_{{r}}\right]\label{eq:eps_H}\\
\epsilon_{\varphi_{1}} & =\frac{1}{\mathcal{D}}\left[\left(\left(-12\,\alpha+24\right){r_{{1}}}^{2}+\left(-6\,{\alpha}^{2}+24\,\alpha-24\right)r_{{1}}\right)r_{{3}}^{4}+\left(48\,\alpha\,{r_{{1}}}^{2}+\left(48\,{\alpha}^{2}-96\,\alpha\right)r_{{1}}\right)r_{{3}}^{3}\right.\nonumber \\
 & \left.+\left(\left(-36\,\alpha-24\right){r_{{1}}}^{2}+\left(-78\,{\alpha}^{2}+72\,\alpha+72\right)r_{{1}}\right)r_{{3}}^{2}+\left(36\,{\alpha}^{2}+24\,\alpha\right)r_{{1}}r_{{3}}\right]r_{{2}}\nonumber \\
 & +\left(\left(4\,\alpha\,\Omega_{{r}}+36\,\alpha-8\,\Omega_{{r}}-72\right)r_{{1}}-2\,{\alpha}^{2}\Omega_{{r}}+6\,{\alpha}^{2}+8\,\alpha\,\Omega_{{r}}-24\,\alpha-8\,\Omega_{{r}}+24\right){r_{{3}}}^{2}\nonumber \\
 & +\left(\left(-16\,\alpha\,\Omega_{{r}}-48\,\alpha\right)r_{{1}}+4\,{\alpha}^{2}\Omega_{{r}}-12\,{\alpha}^{2}-8\,\alpha\,\Omega_{{r}}+24\,\alpha\right)r_{{3}}+\left(12\,\alpha\,\Omega_{{r}}-36\,\alpha+8\,\Omega_{{r}}-24\right)r_{{1}}\nonumber \\
 & -\left.2\,{\alpha}^{2}\Omega_{{r}}+6\,{\alpha}^{2}-8\,\alpha\,\Omega_{{r}}+24\,\alpha-8\,\Omega_{{r}}+24\right]\label{eq:eps_phi1}\\
\epsilon_{\varphi_{2}} & =-\frac{2}{\mathcal{D}}\,\left[\left(-6\,{\alpha}^{2}+24\,\alpha-24\right)r_{{1}}{r_{{3}}}^{4}+\left(30\,{\alpha}^{2}-60\,\alpha\right)r_{{1}}{r_{{3}}}^{3}+\left(-69\,{\alpha}^{2}+12\,\alpha+12\right)r_{{1}}r_{{3}}^{2}\right.\nonumber \\
 & \left.+\left(72\,{\alpha}^{2}+48\,\alpha\right)r_{{1}}r_{{3}}+\left(-27\,{\alpha}^{2}-36\,\alpha-12\right)r_{{1}}\right]r_{{2}}\nonumber \\
 & +2\,\left({\alpha}^{2}\Omega_{{r}}-3\,{\alpha}^{2}-4\,\alpha\,\Omega_{{r}}+12\,\alpha+4\,\Omega_{{r}}-12\right)r_{{3}}^{2}\nonumber \\
 & +2\,\left(-2\,{\alpha}^{2}\Omega_{{r}}+6\,{\alpha}^{2}+4\,\alpha\,\Omega_{{r}}-48\,\alpha\,r_{{1}}-12\,\alpha\right)r_{{3}}\nonumber \\
 & +\left.2\,\left(72\,\alpha+48\right)r_{{1}}+2\,{\alpha}^{2}\Omega_{{r}}-6\,{\alpha}^{2}+8\,\alpha\,\Omega_{{r}}-24\,\alpha+8\,\Omega_{{r}}-24\right]\label{eq:eps_phi2}
\end{flalign}
where
\begin{flalign}
\mathcal{D}= & 2r_{1}\left(8\left(3\alpha-2\alpha r_{3}+2\right)-r_{2}\left(r_{3}-1\right){}^{2}\left(3\alpha-(\alpha-2)r_{3}+2\right){}^{2}\right)\nonumber \\
 & +\left((\alpha+2)^{2}+(\alpha-2)r_{3}\left((\alpha-2)r_{3}-2\alpha\right)\right)\left(r_{2}\left(3\alpha+3(\alpha-2)r_{3}^{2}-6\alpha r_{3}+2\right)-8\right).
\end{flalign}
Now, the cosmological dynamics of the model is studied by taking the
derivative of the variables $r_{i}$, and $\Omega_{r}$ with respect
to $N=\textrm{Ln}\,a.$ Doing so we obtain
\begin{eqnarray}
r_{1}' & = & -(\epsilon_{\varphi_{1}}+\epsilon_{H})r_{1},\label{x'}\\
r_{2}' & = & \left(3\epsilon_{\varphi_{2}}-\epsilon_{H}\right)r_{2},\label{y'}\\
r_{3}' & = & \left(\epsilon_{\varphi_{2}}-\epsilon_{\varphi_{1}}\right)r_{3},\label{z'}\\
\Omega_{r}' & = & -2\left(2+\epsilon_{H}\right)\Omega_{r}.\label{r'}
\end{eqnarray}
For our purpose we just consider Eq.(\ref{z'}) with  the relations (\ref{eq:eps_H})-(\ref{eq:eps_phi2}) to get 
\par\end{flushleft}

\begin{equation}
r_{3}' =  \frac{2}{\mathcal{D}}r_{1}r_{3}\left(r_{3}-1\right)
\left((\alpha-2)r_{3}-3\alpha-2\right)\left(3r_{2}\left(2+3\alpha-6\alpha r_{3}+\left(3\alpha+2r_{1}-6\right)r_{3}^{2}\right)
-2\left(\Omega_{r}+9\right)\right).\label{eq:de3}
\end{equation}
In terms of the variables $r_{i}$ the dark energy density and the
dark energy equation of state are given by
\begin{flalign}
\Omega_{DE} & =-\frac{1}{2}\,r_{{2}}\left(\,r_{{3}}^{2}\left(2\,r_{{1}}+3\alpha-6\right)-6\,r_{{3}}\alpha+3\,\alpha+2\right),\label{eq:DE}\\
w_{DE} & =\frac{3+2\epsilon_{H}+\Omega_{r}}{\Omega_{DE}}.\label{eq:EOSDE}
\end{flalign}
We also define the effective equation of state, $\omega_{eff}=-1-\frac{2}{3}\epsilon_{H}$.
Setting $r_{3}=1$ in (\ref{eq:DE}) it follows that the dark energy
density of the BG model is proportional to that of the cubic SG model,
$\Omega_{DE}^{BG}=r_{1}^{2}\Omega_{DE}^{SG}$. This means that two
models are governed by the same dynamics, and we expect to find a
 tracker solution similar to that of the cubic SG model.

The fixed points $\left(r_{1c},r_{2c},r_{3c},\Omega_{rc}\right)$
of the BG model are solutions of the equations $r_{i}'=0$ and $\Omega_{r}'=0$.
However, we need also to identify the stability of these fixed points
by calculating the matrix of the coefficients of the perturbed equations
to linear order. A critical point is said to be stable only if all
the eigenvalues of the perturbation matrix are negatives, unstable
if the eigenvalues are positives and saddle if the eigenvalues have
different signs. By inspection, Eq. (\ref{eq:de3}) admits two fixed
points given by
\[
r_{3}=1,\quad\textrm{and}\quad r_{3}=\frac{3\alpha+2}{\alpha-2}.
\]
These solutions signal the existence of two tracker solutions. There
is also an other solution given by $r_{1}=0$ which gives the so-called
small regime. In Table.\ref{t1}, we list the fixed points, the conditions
of their existence and their stability in the radiation, matter and
de Sitter epochs. The interesting fact is the emergence of two sets
of fixed points in each cosmological epoch. The first set of fixed
points, independent of the coupling constant $\alpha$, is an extension
of the one found in the context of the SG field model, while the second
set of fixed points is intrinsic to the cubic BG model and may induce
a new behavior of dark energy equation of state, particularly in the
matter dominated epoch.

\begin{table}[h]
\begin{centering}
\begin{tabular}{|l|c|c|c|c|c|c|c|c|c|}
\hline
{\tiny{}Point} & {\tiny{}$r_{1c}$} & {\tiny{}$r_{2c}$} & {\tiny{}$r_{3c}$} & {\tiny{}$\Omega_{r}$} & {\tiny{}Existence} & {\tiny{}eigenvalues} & {\tiny{}Stability} & {\tiny{}$\omega_{eff}$} & {\tiny{}$\Omega_{m}$}\tabularnewline
\hline
\hline
\textbf{\textit{A}} & {\tiny{}$0$} & {\tiny{}$0$} & {\tiny{}$r_{3}$} & {\tiny{}$1$} & {\tiny{}$\left(\left(\alpha+2\right)^{2}-2\left(2-\alpha\right)\alpha r_{3}+\left(2-\alpha\right)^{2}r_{3}^{2}\right)\neq0$} & {\tiny{}$\left(0,0,\frac{5}{2},\frac{1}{2}\right)$} & {\tiny{}Unstable} & {\tiny{}$\frac{1}{3}$} & {\tiny{}$0$}\tabularnewline
\hline
\textbf{\textit{B}} & {\tiny{}$0$} & {\tiny{}$0$} & {\tiny{}$r_{3}$} & {\tiny{}$0$} & {\tiny{}$\left(\left(\alpha+2\right)^{2}-2\left(2-\alpha\right)\alpha r_{3}+\left(2-\alpha\right)^{2}r_{3}^{2}\right)\neq0$} & {\tiny{}$\left(0,0,\frac{9}{4},-\frac{3}{4}\right)$} & {\tiny{}Saddle} & {\tiny{}0} & {\tiny{}$1$}\tabularnewline
\hline
\textbf{\textit{C}} & {\tiny{}$1$} & {\tiny{}$0$} & {\tiny{}$1$} & {\tiny{}$1$} & {\tiny{}$\left(\alpha+2\right)\neq0$} & {\tiny{}$\left(0,8,-5,-5\right)$} & {\tiny{}Saddle} & {\tiny{}$\frac{1}{3}$} & {\tiny{}$0$}\tabularnewline
\hline
\textbf{\textit{D}} & {\tiny{}$1$} & {\tiny{}$0$} & {\tiny{}$1$} & {\tiny{}$0$} & {\tiny{}$\left(\alpha+2\right)\neq0$} & {\tiny{}$\left(0,6,-\frac{9}{2},-\frac{9}{2}\right)$} & {\tiny{}Saddle} & {\tiny{}$0$} & {\tiny{}$1$}\tabularnewline
\hline
\textbf{\textit{T1}} & {\tiny{}$1$} & {\tiny{}$1$} & {\tiny{}$1$} & {\tiny{}$0$} & {\tiny{}$\left(\alpha+2\right)\neq0$} & {\tiny{}$\left(0,-3,-3,-3\right)$} & {\tiny{}Stable} & {\tiny{}$-1$} & {\tiny{}$0$}\tabularnewline
\hline
\textbf{\textit{E}} & {\tiny{}$-\frac{\left(\alpha+1\right)\left(\alpha-2\right)}{3\alpha+2}$} & {\tiny{}$0$} & {\tiny{}$\frac{3\alpha+2}{\alpha-2}$} & {\tiny{}$1$} & {\tiny{}$\left(\alpha+1\right)\left(\alpha+2\right)\left(\alpha-2\right)\left(3\alpha+2\right)\neq0$ } & {\tiny{}$\left(0,8,-5,-5\right)$} & {\tiny{}Saddle} & {\tiny{}$\frac{1}{3}$} & {\tiny{}$0$}\tabularnewline
\hline
\textbf{\textit{F}} & {\tiny{}$-\frac{\left(\alpha+1\right)\left(\alpha-2\right)}{3\alpha+2}$} & {\tiny{}$0$} & {\tiny{}$\frac{3\alpha+2}{\alpha-2}$} & {\tiny{}$0$} & {\tiny{}$\left(\alpha+1\right)\left(\alpha+2\right)\left(\alpha-2\right)\left(3\alpha+2\right)\neq0$} & {\tiny{}$\left(0,6,-\frac{9}{2},-\frac{9}{2}\right)$} & {\tiny{}Saddle} & {\tiny{}0} & {\tiny{}1}\tabularnewline
\hline
\textbf{\textit{T2}} & {\tiny{}$-\frac{\left(\alpha+1\right)\left(\alpha-2\right)}{3\alpha+2}$} & {\tiny{}$-\frac{\alpha-2}{\left(3\alpha+2\right)\left(\alpha+1\right)}$} & {\tiny{}$\frac{3\alpha+2}{\alpha-2}$} & {\tiny{}$0$} & {\tiny{}$\left(\alpha+1\right)\left(\alpha+2\right)\left(\alpha-2\right)\left(3\alpha+2\right)\neq0$} & {\tiny{}$\left(0,-3,-3,-3\right)$} & {\tiny{}Stable} & {\tiny{}-1} & {\tiny{}$0$}\tabularnewline
\hline
\end{tabular}
\par\end{centering}
\vspace{0.5cm}

\centering{}\caption{\label{t1}Fixed points of the BG model, their stability and existence conditions.}
\end{table}

\section{Analysis of the fixed points\label{sec:Analysis-of-the-fixed-points}}

\subsection{Small regime }

The fixed points \textbf{\textit{A}} and \textbf{\textit{B}} are radiation
and matter dominated points and constitute the regime for which $r_{1},r_{2}\ll1$.
A series expansion in $r_{1}$ and $r_{2}$ lead to the simplified
dynamical equations
\begin{eqnarray}
r_{1}' & = & \frac{\left(\Omega_{r}+9\right)}{4}r_{1},\label{eq:small1}\\
r_{2}' & = & \frac{\left(5\Omega_{r}-3\right)}{4}r_{2},\label{eq:small2}\\
r_{3}' & = & \frac{r_{1}r_{3}\left(r_{3}-1\right)\left(\left(\alpha-2\right)r_{3}-3\alpha-2\right)\left(\Omega_{r}+9\right)}{2\left[(\alpha-2)^{2}r_{3}^{2}-2\alpha(\alpha-2)r_{3}+\left(\alpha+2\right)^{2}\right]},\label{eq:small3}\\
\Omega_{r}' & = & \Omega_{r}\left[\left(\Omega_{r}-1\right)+\frac{3}{8}r_{2}\left(\left(\alpha-2\right)r_{3}^{2}-3\alpha\left(2r_{3}-1\right)+2\right)\left(\Omega_{r}-3\right)\right].
\end{eqnarray}
In the small regime the effective EoS and dark energy parameters are
given by
\begin{alignat}{1}
\omega_{eff} & \approx=\frac{\Omega_{r}}{3}+\frac{1}{8}\left[\left(\alpha-2\right)r_{3}^{2}-2\alpha r_{3}+\alpha+\frac{2}{3}\right]\left(\Omega_{r}-3\right)r_{2}\\
\omega_{DE} & \approx-\frac{1}{12}\left(\Omega_{r}-3\right)+\frac{1}{3}\left[\frac{\Omega_{r}+9}{3\left(\alpha-2\right)r_{3}^{2}-6\alpha r_{3}+3\alpha+2}\right]r_{1}r_{3}^{2}\nonumber \\
 & \quad-\frac{1}{96}\left(\Omega_{r}-3\right)\left(3\left(\alpha-2\right)r_{3}^{2}-6\alpha r_{3}+3\alpha+2\right)r_{2}.
\end{alignat}
In the radiation and matter dominated epochs we obtain $\omega_{DE}=1/6$
and $\omega_{DE}=1/4$, respectively. Integration of Eqs.(\ref{eq:small1})
and (\ref{eq:small2}) in the radiation and matter eras give $r_{1}\propto a^{5/2},\:r_{2}\propto a^{1/2}$
and $r_{1}\propto a^{9/4},\:r_{2}\propto a^{-3/4}$, respectively.
Substituting in Hubble parameter $H(t)$ given by (\ref{eq:h_phi1-2})
we obtain, as expected, $H(t)\propto a^{-2}$ ($H(t)\propto a^{-3/2}$)
in the radiation epoch (matter epoch). In the limit of large $r_{3}$
, Eq.(\ref{eq:small3}) is easily integrated and gives $r_{3}\propto\exp\left(\frac{2a^{\sigma}}{\alpha-2}\right)$
where $\sigma=5/2$ ($\sigma=9/4$) in the radiation (matter) epoch.
Translating these results in terms of the field velocity we get $\dot{\varphi}^{1}\propto t^{-1/4},\:\dot{\varphi}^{2}\propto t^{-1/4}\,\exp\left(\frac{2t^{5/4}}{\alpha-2}\right)$
in the radiation epoch, and $\dot{\varphi}^{1}\propto t^{-1/2},\:\dot{\varphi}^{2}\propto t^{-1/2}\,\exp\left(\frac{2t^{3/2}}{\alpha-2}\right)$
in the matter epoch. Hence the evolution $\dot{\varphi}^{1}$ of $\dot{\varphi}^{2}$
is slower than that of the tracker solutions, $\left\{ \dot{\varphi}^{1},\,\dot{\varphi}^{2}\right\} \propto t.$
We note that to maintain $r_{3}$ large in the radiation and matter
epochs we must impose a large initial condition on $r_{3}$ and then
the evolution of the field $\varphi^{1}$ is slower than that of $\varphi^{2}.$

\subsection{de Sitter fixed points }

As we can see from table \ref{t1} we have two stable de Sitter fixed
points \textbf{\textit{T1}} and\textbf{\textit{ T2}}. The fixed point
\textbf{\textit{T1}} is the same as the one already found in the context
of the SG model and discussed extensively in . The second de Sitter
fixed point \textbf{\textit{T2}} is considered as the signature of
the BG model. Assuming that the coordinates of the fixed point \textbf{\textit{I}}
are all positive lead to the following condition on the coupling constant
\begin{equation}
\alpha<-1.\label{eq:cond}
\end{equation}
We show that the Hubble parameter and field velocity in dS epoch are
given by
\begin{gather}
H_{ds}^{\left(T1\right)}=1,\quad\dot{\varphi}_{ds}^{1\left(T1\right)}=\dot{\varphi}_{ds}^{2\left(T1\right)}=1,\label{eq:DS1}\\
H_{ds}^{\left(T2\right)}=\left|\frac{(3\alpha+2)}{(\alpha-2)(\alpha+1)^{1/2}}\right|,\quad\dot{\varphi}_{ds}^{1\left(T2\right)}=\left|\frac{1}{\left(\alpha+1\right)^{1/2}}\right|,\quad\dot{\varphi}_{ds}^{2\left(T2\right)}=H_{ds}^{\left(T2\right)}.\label{eq:DS2}
\end{gather}
This implies that during the de Sitter epoch ( \textbf{\textit{T2}}
) the variation of the field $\varphi^{2}$ is slower than that of
the field $\varphi^{1}$ for $-2<\alpha<-1$.

From the definition of $r_{1}$ and $r_{2}$ in (\ref{variable}),
the dS fixed points allow for tracker solutions such that $\dot{\varphi}_{I}^{\left(j\right)}H=C_{I}^{(j)}$
where $I=1,\,2$ , $j=\boldsymbol{T1},\,\boldsymbol{T2}$ and $C_{I}^{(J)}$
are constants which can be determined from (\ref{eq:DS1}) and (\ref{eq:DS2}).

Let us consider the radiation and matter dominated epochs ( \textbf{\textit{C}},
\textbf{\textit{D}}, \textbf{\textit{E}}, and \textbf{\textit{F}}
fixed points) and expand the dynamical equations to first order in
$r_{2}$ to obtain along the tracker $\boldsymbol{T1}$
\begin{flalign}
r_{2}'= & \frac{2r_{2}(\Omega_{r}-3r_{2}+3)}{r_{2}+1}\label{eq:t11}\\
\Omega_{r}'= & \frac{\Omega_{r}(\Omega_{r}-7r_{2}-1)}{r_{2}+1}\label{eq:t12}
\end{flalign}
and
\begin{flalign}
r_{2}'= & -\frac{2r_{2}\left((\alpha-2)\Omega_{r}+3\left(\alpha+\left(3\alpha^{2}+5\alpha+2\right)r_{2}-2\right)\right)}{-\alpha+\left(3\alpha^{2}+5\alpha+2\right)r_{2}+2}\label{eq:t21}\\
\Omega_{r}'= & -\frac{\Omega_{r}\left(-\alpha+(\alpha-2)\Omega_{r}+7\left(3\alpha^{2}+5\alpha+2\right)r_{2}+2\right)}{\left(-\alpha+\left(3\alpha^{2}+5\alpha+2\right)r_{2}+2\right)}\label{eq:t22}
\end{flalign}
for the tracker $\boldsymbol{T2}$.

We note that the evolution of $r_{2}$ and $\Omega_{r}$ along the
tracker solution \textbf{\textit{T2}} is a function of the coupling
constant $\alpha$, while along the tracker \textbf{\textit{T1}} we
have exactly the evolution equations of Ref. \cite{obesr-cov-gal}
. The set of equations (\ref{eq:t11}-\ref{eq:t12}) and (\ref{eq:t21}-\ref{eq:t22})
can be written in compact form

\begin{eqnarray}
r_{2}' & = & \frac{2r_{2}\left(\Omega_{r}+3-3\Omega_{DE}^{(i)}\right)}{\Omega_{DE}^{(i)}+1},\label{eq:trac11}\\
\Omega_{r}' & = & \frac{\Omega_{r}\left(\Omega_{r}-7\Omega_{DE}^{T}-1\right)}{\Omega_{DE}^{(i)}+1}.\label{eq:trac12}
\end{eqnarray}
where
\begin{equation}
\Omega_{DE}^{(i)}=\frac{r_{2}}{r_{2c}^{(i)}}.\label{eq:trac2DE}
\end{equation}
and $r_{2c}^{(\boldsymbol{T1})}=1$ and $r_{2c}^{(\boldsymbol{T2})}=-\frac{\alpha-2}{\left(3\alpha+2\right)\left(\alpha+1\right)}.$
As long as $\alpha<-1$, $\Omega_{DE}^{(\boldsymbol{I)}}$ remains
positive. In dS epoch we have $\Omega_{DE}=1$ along the two trackers.

Combining Eqs.(\ref{eq:trac11}) and (\ref{eq:trac12}), we show that
\begin{equation}
\frac{r_{2}'}{2r_{2}}-\frac{\Omega_{r}'}{\Omega_{r}}=4.\label{eqr2 and r}
\end{equation}
The integration of this equation with respect to $N$ gives ($a=e^{N}):$
\begin{equation}
r_{2}=d^{(i)}a^{8}\Omega_{r}^{2}.\label{r2=00003Domegar}
\end{equation}
where $d^{(i)}$ is a constant given by $d^{(\boldsymbol{T1})}=\frac{1-\Omega_{m}^{(0)}-\Omega_{r}^{(0)}}{\left(\Omega_{r}^{(0)}\right)^{2}}$,
and $d^{(\boldsymbol{T2})}=-\frac{\alpha-2}{\left(3\alpha+2\right)\left(\alpha+1\right)}d^{(\boldsymbol{T1})}$.
Using Eq.(\ref{r2=00003Domegar}) and $\Omega_{r}=\left(H_{0}/H\right)^{2}\Omega_{r}^{(0)}a^{-4}$,
$\Omega_{m}=\left(H_{0}/H\right)^{2}\Omega_{m}^{(0)}a^{-3}$ in (\ref{eq:trac12}),
and solving for $\Omega_{r}$, we finally obtain the Hubble parameter
along both the trackers
\begin{equation}
\left(\frac{H}{H_{0}}\right)^{2}=\frac{1}{2}\Omega_{m}^{(0)}(z+1)^{3}+\frac{1}{2}\Omega_{r}^{(0)}(z+1)^{4}+\sqrt{1-\Omega_{m}^{(0)}-\Omega_{r}^{(0)}+\left(\Omega_{m}^{(0)}(z+1)^{3}+\Omega_{r}^{(0)}(z+1)^{4}\right){}^{2}}.\label{H}
\end{equation}
This equation does not show any dependence on the coupling $\alpha$,
and is exactly the one obtained on the SG field model Ref.
\cite{obesr-cov-gal}. On the other hand, the effective equation of
state $w_{eff}$ and dark energy equation of state $w_{DE}$
on the tracker solutions are given by

\begin{equation}
w_{eff}^{(j)}=\frac{\Omega_{r}-6\Omega_{DE}^{(j)}}{3\left(1+\Omega_{DE}^{(j)}\right)}\quad,w_{DE}^{(j)}=-\frac{\Omega_{r}+6}{3\left(1+\Omega_{DE}^{(j)}\right)}.
\end{equation}
In the early cosmological epoch in which $\Omega_{DE}^{(i)}\ll1$
these relations reduce to $w_{eff}\simeq\Omega_{r}/3$ and $w_{DE}\simeq-2-\Omega_{r}/3.$
Then, in the radiation epoch ($\Omega_{DE}^{(j)}\ll1$ and $\Omega_{r}\approx1$)
we have $w_{eff}\simeq1/3$ and $w_{DE}\simeq-7/3,$ while in matter
epoch ($\Omega_{DE}^{(j)}\ll1$ and $\Omega_{r}\ll1$) we have $w_{eff}\simeq0$
and $w_{DE}\simeq-2.$ In dS epoch we obtain $w_{eff}=-1$ and $w_{DE}=-1.$
Although the dynamical evolution of $r_{2}$ and $\Omega_{r}$ is
different along the trackers $\boldsymbol{T1}$ and $\boldsymbol{T2}$,
we found that the evolution of the dark energy equation of state along
the tracker solutions of the BG field model is identical to that of
SG field model, and then is also plagued by the same bad behavior
in the matter epoch, where it reaches the value $w_{DE}=-2$. Hence,
the tracker solution of the BG model is in tension with cosmological
data with respect to the $\Lambda\textrm{CDM}$ model \cite{obesr-cov-gal}.

\subsection{Dark energy solution\label{subsec:Dark-energy-solution}}

The dynamical variable $r_{3}$ allows us to investigate deeply the
competition between the fields of the BG model since this variable
controls the rate of the evolution of the field $\varphi^{2}$ with
to the field $\varphi^{1}$. Indeed, besides the solutions listed
in table. \ref{t1}, the dynamical equations exhibit a rich dark energy
structure in the case where $r_{3}$ dominates over $r_{1}$ and $r_{2}$
in the radiation and matter epochs. We also choose $r_{2}$ much smaller
than $1$ to maintain $\Omega_{DE}\ll1$ in these epochs. In this
regime the dynamical system reduce to
\begin{flalign}
r_{1}'\approx & \frac{1}{4}\,\frac{r_{1}\left(\Omega_{r}+9\right)\left(\alpha+2\,r_{1}-2\right)}{\alpha-2},\label{eq:dark_eq1}\\
r'_{3}\approx & \frac{\left(\Omega_{r}+9\right)r_{1}r_{3}}{2\left(\alpha-2\right)},\label{eq:dark_eq2}\\
\Omega'_{r}\approx & \Omega_{r}\left(\Omega_{r}-1\right).
\end{flalign}
Integrating Eqs.(\ref{eq:dark_eq1}) and (\ref{eq:dark_eq2}) we obtain
\begin{equation}
r_{1}\left(a\right)=\frac{\left(\alpha-2\right)a^{\sigma}}{B\left(\alpha-2\right)-2a^{\sigma}},\quad r_{3}\left(a\right)=\frac{A}{2a^{\sigma}-B\left(\alpha-2\right)}
\end{equation}
where $A$ and $B$ are constants of integration and $\sigma=5/2$,
$\sigma=9/4$ for $\Omega_{r}=1$ and $\Omega_{r}=0,$ respectively.
It is clear that $r_{1}=\left(2-\alpha\right)/2$ is one solution
of Eq.(\ref{eq:dark_eq1}). We have to choose $B\approx0$ and $A\approx1$
to maintain the dominance of $r_{3}$ over $r_{1}$ and $r_{2}$.
Then, we are left with the following new solution $\boldsymbol{J}_{\Omega_{r}}=\left(r_{1}=\left(2-\alpha\right)/2,\:r_{2}\approx0,\:r_{3}\gg1,\:\Omega_{r}\right).$
Along this solution the effective and dark energy equations of state
do not dependent on the coupling constant $\alpha$ and are given
by
\begin{equation}
w_{eff}\approx\frac{\Omega_{r}}{3},\quad w_{DE}\approx-\frac{1}{2}-\frac{\Omega_{r}}{6}.\label{eq:omega_large}
\end{equation}

This approximate solution (\ref{eq:omega_large}) is not accurate
in dS epoch. The behavior of the $w_{DE}$ along the dark energy solution
is slightly improved compared to that along the tracker solutions
$\boldsymbol{T1}$ and $\boldsymbol{T2}$. In fact, during radiation
and matter dominated epochs we have $w_{DE}=-2/3$ and $w_{DE}=-1/2$,
respectively. This means that the cosmological dynamics with initial
conditions $r_{3}^{(s)}$ much larger than $1$ escape the tracker
curves with $w_{DE}=-2$ during matter dominated epoch. In terms of
the field velocity we obtain $H\dot{\varphi}^{1}=\frac{B\left(\alpha-2\right)-2a^{\sigma}}{\left(\alpha-2\right)a^{\sigma}}$
and $H\dot{\varphi}^{2}=\frac{A}{\left(\alpha-2\right)a^{\sigma}}$.
For $B\approx0$ we get $H\dot{\varphi}^{1}=\textrm{constant}$. We
then obtain $\dot{\varphi}^{1}\propto t$, $\dot{\varphi}^{2}\propto t^{9/4}$
and $\dot{\varphi}^{1}\propto t,$ $\dot{\varphi}^{2}\propto t^{5/2}$
in radiation and matter epochs, respectively. These behaviors show
the dominance of the field $\varphi^{2}$ over $\varphi^{1}$ during
these epochs. A second solution to Eqs. (\ref{eq:dark_eq1}) and (\ref{eq:dark_eq2})
is $r_{1}=0,\:r_{2}=0$ which is already given by the fixed points
\textbf{\textit{A}} and \textbf{\textit{B}}, and for which we have
$w_{eff}\approx\frac{\Omega_{r}}{3},\quad w_{DE}\approx\frac{1}{4}-\frac{\Omega_{r}}{12}.$

\begin{figure}
\begin{centering}
\includegraphics[width=10cm,height=6cm]{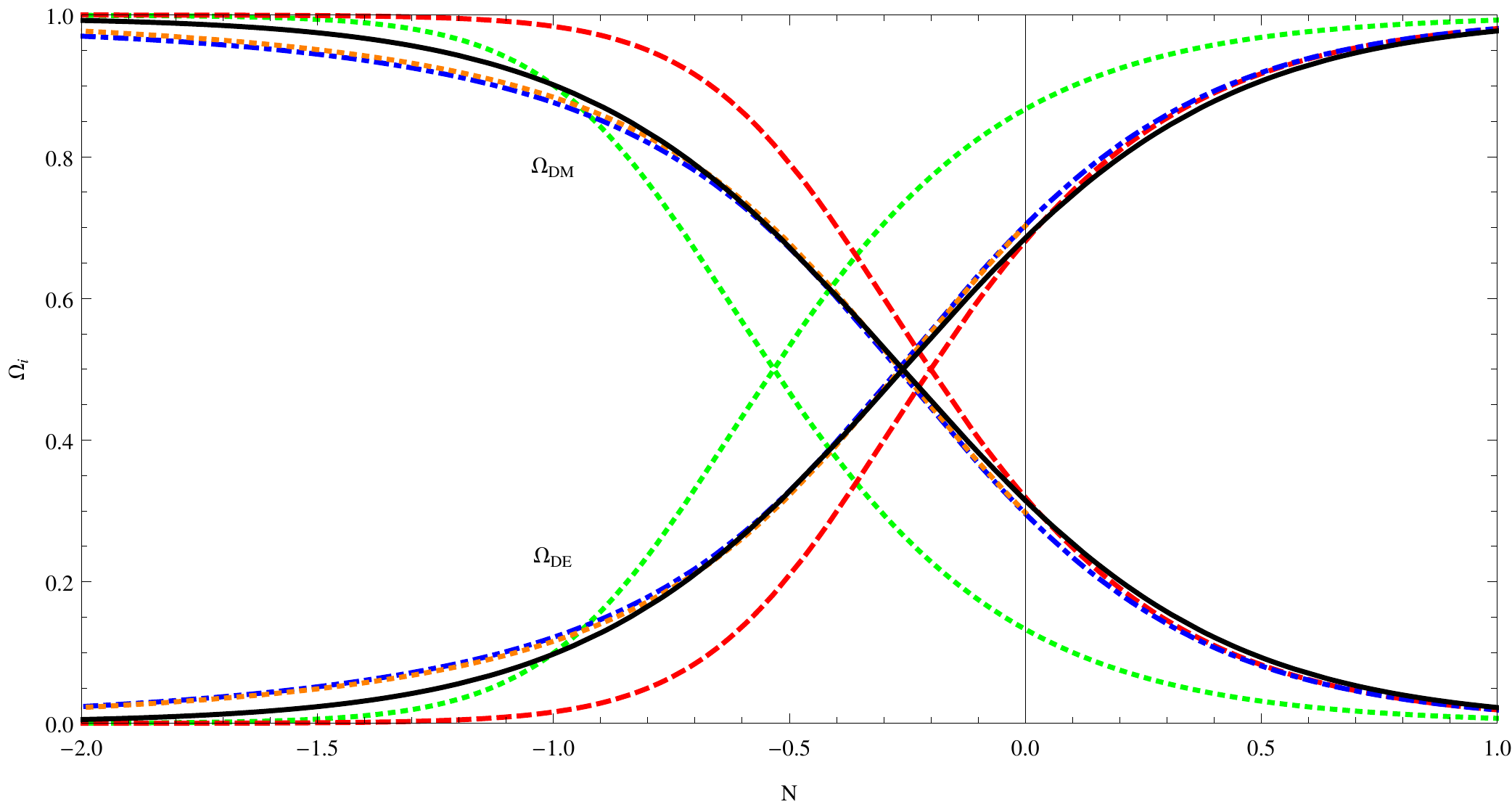}
\caption{\label{fig10}Evolution of energy densities for BG and $\Lambda\textrm{CDM}$
models. For all the plots we take $\alpha=-19.8$ and $\Omega_{m}^{(0)}h^{2}=0.1426$
(Planck 2015: TT+lowE). The initial conditions for BG1 model are: $r_{1}^{(s)}=5.5\times10^{-14},\:r_{2}^{(s)}=1.348\times10^{-22}$
and $r_{3}^{(s)}=3\times10^{6}$ (Blue), and   $r_{1}^{(s)}=\left(2-\alpha\right)/2,\:r_{2}^{(s)}=5\times10^{-62}$, $r_{3}^{(s)}=2.4\times10^{20}$ (Orange) for BG2 model. 
The curves for tracker
solutions $\boldsymbol{T1}$ and $\boldsymbol{T2}$ and $\Lambda\textrm{CDM}$
model are shown by green, red and black colors, respectively.}
\end{centering}
\end{figure}

\begin{figure}
\begin{minipage}[b]{0.4\textwidth}
\includegraphics[width=7cm,height=6cm]{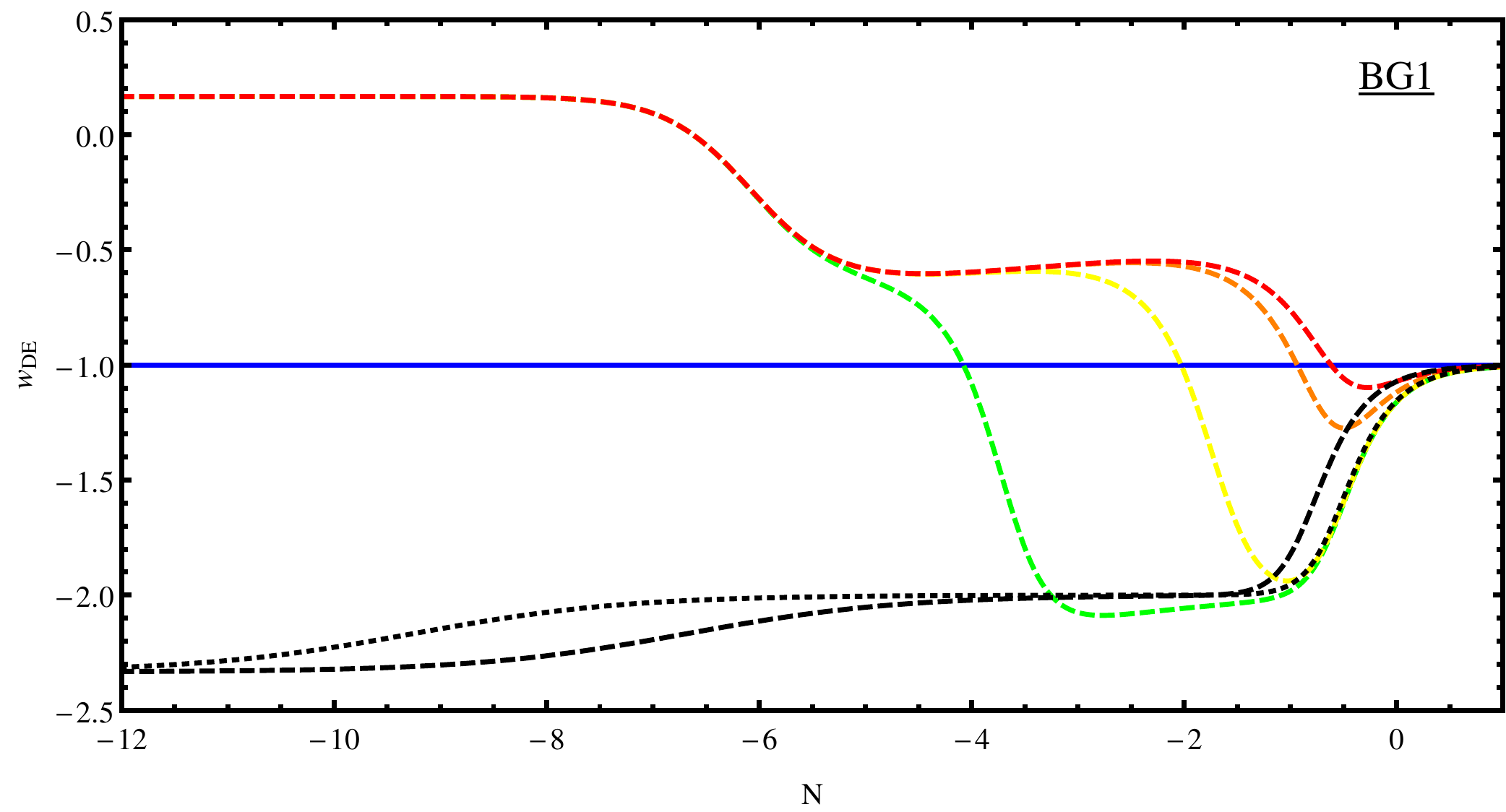}
\end{minipage}
\hspace*{2cm}
 \begin{minipage}[b]{0.4\textwidth}
\includegraphics[width=7cm,height=6cm]{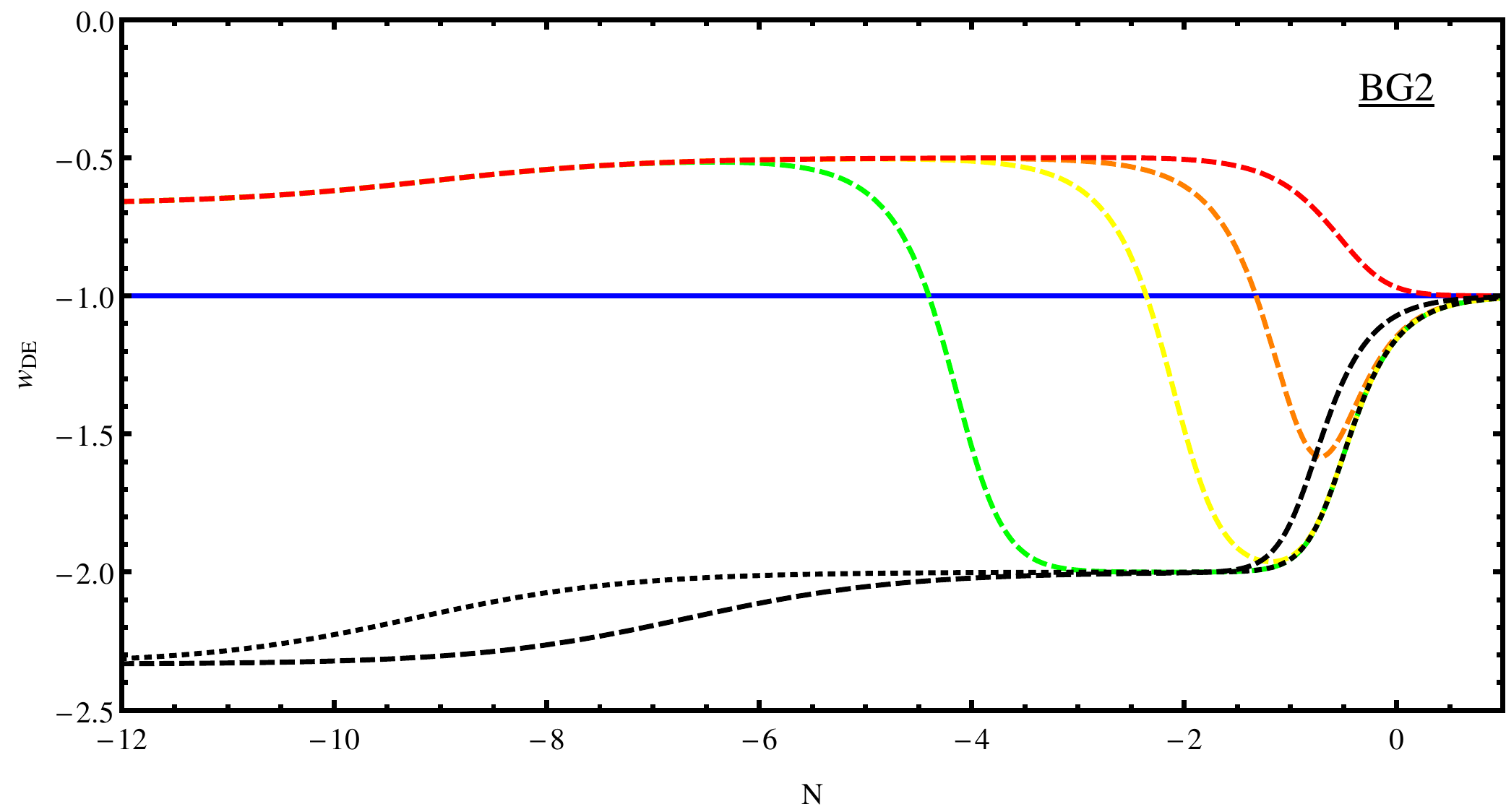}
\end{minipage}
\caption{\label{fig11}Evolution of dark energy equation of state $w_{DE}$
for BG1 and BG2 models. The initial conditions for BG1 models are:
$r_{1}^{(s)}=10^{-14},\:r_{2}^{(s)}=2\times10^{-23}$ and $r_{3}^{(s)}=10^{3}$
(Green), $r_{3}^{(s)}=10^{5}$ (Yellow), $r_{3}^{(s)}=2\times10^{6}$(Orange),
$r_{3}^{(s)}=5\times10^{6}$(Red). The initial conditions for BG2
model are: $r_{1}^{(s)}=10.0522,\:r_{2}^{(s)}=2\times10^{-63}$ and
$r_{3}^{(s)}=10^{17}$ (Green), $r_{3}^{(s)}=10^{19}$ (Yellow), $r_{3}^{(s)}=2\times10^{20}$
(Orange), $r_{3}^{(s)}=5\times10^{20}$(Red). The tracker solutions
$\boldsymbol{T1}$ and $\boldsymbol{T2}$ are shown by the black dashed
and dotted curves, respectively.}

\end{figure}

We proceed now to a numerical integration of equations of the BG model
considering two cases. The first case, labeled BG1 model, corresponds
to the integration of the full set of dynamical equations (\ref{x'}-\ref{r'}),
and the second case, labeled BG2 model, is based on the integration
of the dark energy solution found in Sec. \ref{subsec:Dark-energy-solution}.
In Fig. \ref{fig10}, we plot the evolution of the dimensionless energy
densities for the BG1 and BG2 models with that of the tracker solutions
and compare with the energy density evolution in the $\Lambda\textrm{CDM}$
model. We start the simulation at early times in the deep radiation
epoch at $N^{(s)}=-20\:\left(z^{(s)}\approx4.85\times10^{8}\right)$.
The initial conditions on $r_{i}^{(s)}$ are chosen such that the
today values of the energy densities are compatible with the Planck
2015 data (TT+lowE), $\Omega_{m}=0.315$ and $\Omega_{DE}=0.685.$
We observe that the evolution in the BG models is compatible with
that of the $\Lambda\textrm{CDM }$model, whereas the evolution of
the energy densities obtained from the tracker solutions is not, particularly
the tracker solution $\boldsymbol{T1}$. An other observation is that
the BG models follow the tracker solution $\boldsymbol{T2}$ earlier
or later depending on the initial conditions. In fact, this approach
to the tracker $\boldsymbol{T2}$ is best seen in Fig. \ref{fig11}
where we show the evolution of the dark energy parameter of state
$w_{DE}$. Indeed, $w_{DE}$ follows the tracker curve early at moderate
initial condition $r_{3}^{(s)}$ for the BG1 and BG2 models. The difference
between the two models occurs at around $a<6\times10^{-3}$ where
the $w_{DE}$ follow two different paths in the radiation epoch and
at the onset of the matter epoch. For $a>6\times10^{-3}$, $w_{DE}$
reaches the value $-1/2$, as predicted by the dark energy solution
(\ref{eq:omega_large}), before decreasing to values around $-1$
at the onset of the dS epoch. We observe also that a large initial
condition on $r_{3}$ is the best realization for $w_{DE}$ which
becomes very close to $w_{DE}=-1.$ Finally, we note that large initial
conditions $r_{3}^{(s)}$ does not necessarily imply large present-day
values on $r_{i}$, as we obtain for the red curve shown in Fig. \ref{fig11}
the following values: $r_{1}^{(0)}=6.0185,\:r_{2}^{(0)}=0.0125$ and
$r_{3}^{(0)}=2.9824.$ We conclude that the tracker solutions are
incompatible with $\Lambda\textrm{CDM}$ model in the matter dominated
epoch, and that the evolution of $w_{DE}$ calculated from the dark
energy solution in the regime of $r_{3}^{(s)}>>\left\{ 1,\:r_{1}^{(s)}\right\} $
and $r_{2}^{(s)}\ll1$ prevents the approach to the bad behavior of
the tracker in the matter epoch. A similar behavior has been recently
obtained with the Galileon ghost condensate model \cite{Galileon-condensate,condensate-gal}.

\section{Growth rate of matter perturbations\label{sec:Growth-rate-of}}

The study of the growth rate of cosmological density perturbations
has become a powerful tool to distinguish between cosmological models
based on modified theories of gravity  and dark energy models. Even
all models can perfectly mimic the $\Lambda$CDM evolution at the
background level, they all intrinsically alter the structure formation.
An important probe in this context is the evolution of linear matter
density contrast $\delta_{m}\equiv\delta\rho_{m}/\rho_{m}$ which
verify the following equation
\begin{equation}
\ddot{\delta}+2H\dot{\delta}-4\pi G_{eff}\rho_{m}\approx0\label{eq:delta}
\end{equation}
where $G_{eff}$ is a function of the scale factor and the cosmological
scale. The matter density contrast is related to the observed quantity
$f\left(a\right)\sigma_{8}\left(a\right)$ where $f\left(a\right)=d\ln\delta\left(a\right)/d\ln\left(a\right)$
and $\sigma_{8}\left(a\right)=\sigma_{8}\delta_{m}\left(a\right)/\delta_{m}\left(1\right)$
is the rms fluctuations of the linear density field inside a radius
of $8h^{-1}\textrm{Mpc, and \ensuremath{\sigma_{8}} is its present value. }$

We propose now to study the evolution of the equations (\ref{EE}),
(\ref{energy-tensor-moment}) and (\ref{eq:contunu}) at the perturbed
level. We consider only scalar perturbations of the flat FRW metric
in the Newtonian gauge
\begin{equation}
ds^{2}=(1-2\Phi)dt^{2}-a\left(t\right)^{2}(1+2\Psi)\delta_{ij}dx^{j}dx^{i},\label{ds-per}
\end{equation}
where $\Phi$ and $\Psi$ are scalar metric perturbations related
to the Newtonian potential and perturbation of the spatial three-curvature.
Perturbing the scalar fields and the matter density, $\varphi^{I}\rightarrow\varphi^{I}\left(t\right)+\delta\varphi^{I}\left(t,\,x^{i}\right)\,\rho_{m}\rightarrow\rho_{m}\left(1+\delta\left(t,\,x^{i}\right)\right)$
and keeping perturbations at first order, the Einstein equations (\ref{EE})
thus take the form

$(0,\,0)$:
\begin{eqnarray}
\frac{k^{2}}{a^{2}}\left(\Psi-\frac{1}{2}b_{IJK}\dot{\varphi}^{J}\dot{\varphi}^{K}\delta\varphi^{I}\right)-\frac{1}{4}a_{IJ}\delta\left[\dot{\varphi}^{J}\dot{\varphi}^{I}\right]+9Hb_{IJK}\dot{\varphi}^{I}\dot{\varphi}^{J}\delta\dot{\varphi}^{K}\nonumber \\
+3\left(H-\frac{1}{2}b_{IJK}\dot{\varphi}^{I}\dot{\varphi}^{J}\dot{\varphi}^{K}\right)\dot{\Psi}-\left(3Hb_{IJK}\dot{\varphi}^{I}\dot{\varphi}^{J}\dot{\varphi}^{K}+2\rho_{m}\right)\Phi & = & \rho_{m}\delta,\label{e00}
\end{eqnarray}

$(i,\,i)$:
\begin{eqnarray}
-\frac{k^{2}}{a^{2}}\left(\Psi-\Phi\right)-3\left(\ddot{\Psi}+3H\dot{\Psi}\right)-\frac{3}{4}a_{IJ}\delta\left(\dot{\varphi}^{J}\dot{\varphi}^{I}\right)-3b_{IJK}\dot{\varphi}^{I}\dot{\varphi}^{J}\delta\dot{\varphi}^{K}-\frac{3}{2}b_{IJK}\dot{\varphi}^{J}\dot{\varphi}^{I}\delta\ddot{\varphi}^{K}\nonumber \\
-3\left(H+\frac{1}{2}b_{IJK}\dot{\varphi}^{J}\dot{\varphi}^{I}\dot{\varphi}^{K}\right)\dot{\Phi}-\frac{3}{2}\left(6\dot{H}+4H^{2}+a_{IJ}\dot{\varphi}^{J}\dot{\varphi}^{I}+2b_{IJK}\dot{\varphi}^{I}\dot{\varphi}^{J}\ddot{\varphi}^{K}\right)\left(\Phi+\Psi\right) & = & 0,\label{eii}
\end{eqnarray}

$(i,\,0)$:
\begin{eqnarray}
-\dot{\Psi}-\frac{1}{2}b_{IJK}\dot{\varphi}^{I}\dot{\varphi}^{J}\delta\dot{\varphi}^{K}-\left(\frac{1}{2}a_{IJ}\dot{\varphi}^{J}-\frac{3}{2}b_{IJK}\dot{\varphi}^{J}\dot{\varphi}^{K}\right)\delta\varphi^{I}-\left(H+b_{IJK}\dot{\varphi}^{I}\dot{\varphi}^{J}\dot{\varphi}^{K}\right)\Phi & = & \rho_{m}v,\label{ei0}
\end{eqnarray}

$(i\neq j)$:
\begin{equation}
\partial_{i}\partial_{j}\left(\Psi-\Phi\right)=0,\label{eij}
\end{equation}
where $k$ is the cosmological scale.

The BG field equations (\ref{eq:BG_f1}), up to linear order in perturbations,
are given by

\begin{eqnarray}
\left(a_{IJ}+6b_{IJK}H\dot{\varphi}^{K}\right)\delta\ddot{\varphi}^{J}+3\left(a_{IJ}H+2b_{JIK}(H\ddot{\varphi}^{K}+3H^{2}\dot{\varphi}^{K}+\dot{H}\dot{\varphi}^{K})\right)\delta\dot{\varphi}^{J}\nonumber \\
+\frac{k^{2}}{a^{2}}\left(a_{IJ}-2b_{IJK}\left(2H\dot{\varphi}^{K}+\ddot{\varphi}^{K}\right)\right)\delta\varphi^{J}\nonumber \\
+3b_{IJK}\dot{\varphi}^{J}\dot{\varphi}^{K}\ddot{\Phi}-\left(2a_{IJ}\dot{\varphi}^{J}+9Hb_{IJK}\dot{\varphi}^{J}\dot{\varphi}^{K}+6b_{IJK}\dot{\varphi}^{J}\ddot{\varphi}^{K}\dot{\Phi}\right)\dot{\Phi}\nonumber \\
-2\left(a_{IJ}H\dot{\varphi}^{J}+4b_{IJK}H\dot{\varphi}^{J}\ddot{\varphi}^{K}+6b_{IJK}\dot{H}\dot{\varphi}^{J}\dot{\varphi}^{K}+18H^{2}b_{IJK}\dot{\varphi}^{J}\dot{\varphi}^{K}-\frac{k^{2}}{2a^{2}}b_{IJK}\dot{\varphi}^{J}\dot{\varphi}^{K}\right)\Phi & = & 0.\label{eqfieldper}
\end{eqnarray}
In deriving Eq. (\ref{eqfieldper}) we have used the background equation
of motion (\ref{FRW-scalarfield}) and $\Psi=\Phi$ from (\ref{eij}).
Similarly, the perturbed equations of motion for pressurless matter
field (\ref{eq:contunu}) are given by
\begin{flalign}
\dot{\delta\rho_{m}}+3H\delta\rho_{m}=\left(\frac{k}{a^{2}}\right) & \rho_{m}v-3\rho_{m}\dot{\Phi},\label{eq:mattper}\\
\dot{v} & =\Phi\label{eq:eqvper}
\end{flalign}
 where $v$ is the potential of velocity matter perturbation.

Defining the gauge-invariant matter density contrast
\begin{equation}
\delta_{m}\coloneqq\delta_{m}-3Hv
\end{equation}
we write the matter field perturbation in Fourier space as
\begin{equation}
\ddot{\delta}_{m}+2H\dot{\delta}_{m}+\frac{k^{2}}{a^{2}}\Phi=3\left(\ddot{Q}+2H\dot{Q}\right)\label{edeltam}
\end{equation}
where $Q=Hv-\Phi$. Since matter perturbations evolve on spatial scales
much smaller than of the Hubble horizon ($k\gg aH$), we use the so
called quasi-static approximation on sub-horizon scales. Under this
approximation, the dominant terms in the perturbed equations are those
including $\delta_{m}$ and $k^{2}/a^{2}$ . Then, Eqs. (\ref{eij}),
(\ref{e00}), (\ref{edeltam}) and (\ref{eqfieldper}) in the sub-horizon
approximation read
\begin{gather}
\frac{k^{2}}{a^{2}}\left(\Phi-\frac{1}{2}b_{IJK}\dot{\varphi}^{J}\dot{\varphi}^{K}\delta\varphi^{I}\right)+\rho_{m}\delta_{m}=0\label{eq:sub-hor00}\\
b_{IJK}\dot{\varphi}^{J}\dot{\varphi}^{K}\Phi+\left(a_{IJ}-2b_{IJK}\left(2H\dot{\varphi}^{K}+\ddot{\varphi}^{K}\right)\right)\delta\varphi^{J}=0\label{eq:sub-horifield}\\
\ddot{\delta}_{m}+2H\dot{\delta}_{m}+\frac{k^{2}}{a^{2}}\Phi=0\label{eq:eqdelta}
\end{gather}
These equations can be solved for $\Phi$, $\delta\varphi^{1}$ and
$\delta\varphi^{2}$, and as a result we obtain
\begin{eqnarray}
-\frac{k^{2}}{a^{2}}\Phi & = & \frac{\textrm{Det \ensuremath{\left[\textrm{K}\right]}}}{\textrm{Det}\left[\textrm{M}\right]}\rho_{m}\delta_{m},\label{eqpoison}\\
-\frac{k^{2}}{a^{2}}\delta\varphi^{I} & = & \frac{\textrm{L}_{2}\textrm{K}_{1I}-\textrm{L}_{1}\textrm{K}_{2I}}{\textrm{Det}\left[\textrm{M}\right]}\rho_{m}\delta_{m},\label{eqpoisson2}
\end{eqnarray}
where the matrices $\textrm{M}$ and $\textrm{K}$, and the vector
$\textrm{L}$ are given by
\begin{eqnarray}
\textrm{M} & = & \left(\begin{array}{cc}
1\: & \frac{1}{2}\textrm{L}^{T}\\
\textrm{L} & \textrm{K}
\end{array}\right),\label{M}\\
\textrm{K}_{IJ} & = & \left(a_{IJ}-2b_{IJK}\left(2H\dot{\varphi}^{K}+\ddot{\varphi}^{K}\right)\right),\label{K}\\
\textrm{L}_{I} & = & b_{IJK}\dot{\varphi}^{J}\dot{\varphi}^{K}.
\end{eqnarray}
Equation (\ref{eqpoison}) is the modified Poisson equation, $-\frac{k^{2}}{a^{2}}\Phi=G_{eff}\rho_{m}\delta_{m}$,
where the effective gravitational coupling is given by
\begin{equation}
G_{eff}=\frac{\textrm{Det \ensuremath{\left[\textrm{K}\right]}}}{\textrm{Det}\left[\textrm{M}\right]}G_{N},\label{Geff}
\end{equation}
where we have restored Newton's constant $G_{N}$. As we see, the
effective gravitational coupling is a function of $\dot{\varphi}^{K}$
and $\ddot{\varphi}^{K}$, and is therefore subject to change. In
terms of the dynamical variables (\ref{variable}) $G_{eff}$ is expressed
as
\begin{flalign}
\frac{G_{eff}}{G_{N}} & =\frac{8}{D}\left(-(\alpha+2)^{2}r_{3}^{2}\left(\epsilon_{\varphi_{2}}+2\right){}^{2}-\left(\epsilon_{\varphi_{1}}+2\right)\left((\alpha-2)^{2}\epsilon_{\varphi_{1}}+4((\alpha-2)\alpha+2)\right)\right.\nonumber \\
 & +\alpha(\alpha+2)r_{3}\left(3\epsilon_{\varphi_{1}}+\left(2\epsilon_{\varphi_{1}}+5\right)\epsilon_{\varphi_{2}}+8\right)+6r_{1}\Bigl(-(3\alpha-2)\left(\epsilon_{\varphi_{1}}+2\right)+\alpha r_{3}\left(2\epsilon_{\varphi_{2}}+5\right)\Bigr)\Bigr)\label{Geff-1}
\end{flalign}
where
\begin{eqnarray}
D & = & r_{2}\left((\alpha+2)^{3}r_{3}^{4}\left(\epsilon_{\varphi_{1}}-4\epsilon_{\varphi_{2}}-6\right)-(3\alpha-2)\left(4\left(\alpha^{2}-2\alpha+2\right)+(\alpha-2)^{2}\epsilon_{\varphi_{1}}\right)\right.\nonumber \\
 &  & +2\alpha(\alpha+2)^{2}r_{3}^{3}\left(6\epsilon_{\varphi_{2}}+11\right)+2\alpha r_{3}\left(12-12\alpha+17\alpha^{2}+4(\alpha-2)^{2}\epsilon_{\varphi_{1}}\right.\nonumber \\
 &  & \left.+2\left(\alpha^{2}+4\alpha-4\right)\epsilon_{\varphi_{2}}\right)-2\alpha+2r_{3}^{2}\left(8-8\alpha+19\alpha^{2}+3(\alpha-2)^{2}\epsilon_{\varphi_{1}}-(8\right.\nonumber \\
 &  & \left.\left.\left.6\alpha^{2}+8\alpha\right)\epsilon_{\varphi_{2}}\right)\right)+8\left(\left(\epsilon_{\varphi_{1}}+2\right)\left(4\left(\alpha^{2}-2\alpha+2\right)+(\alpha-2)^{2}\epsilon_{\varphi_{1}}\right)+(2\right.\nonumber \\
 &  & \alpha)^{2}r_{3}^{2}\left(\epsilon_{\varphi_{2}}+2\right){}^{2}-\alpha(\alpha+2)r_{3}\left(5\epsilon_{\varphi_{2}}+\epsilon_{\varphi_{1}}\left(2\epsilon_{\varphi_{2}}+3\right)+8\right)\Bigr)-6r_{1}\left(r_{2}(-1\right.\nonumber \\
 &  & \left.+r_{3})^{2}\left(-3\alpha+(\alpha+2)r_{3}+2\right){}^{2}8\left(\alpha r_{3}\left(2\epsilon_{\varphi_{2}}+5\right)-(3\alpha-2)\left(\epsilon_{\varphi_{1}}+2\right)\right)\right).\label{eqdetm0}
\end{eqnarray}
where $\epsilon_{\varphi_{1}},\,\epsilon_{\varphi_{2}}$ are given
by (\ref{eq:eps_phi1}) and (\ref{eq:eps_phi2}). Along the trackers
the effective gravitational coupling simplifies to
\begin{equation}
\frac{G_{eff}}{G_{N}}=\frac{2\left(\Omega_{DE}^{(i)}-2\right)+\Omega_{r}}{4-\Omega_{DE}^{(i)}\left(\Omega_{DE}^{(i)}+3\right)+\Omega_{r}}
\end{equation}
where $\Omega_{DE}^{(i)}$ is given by (\ref{eq:trac2DE}). In radiation and matter epochs, where $\Omega_{DE}^{(i)}\ll1,$
we have $G_{eff}\rightarrow1,$ while In dS epoch where $\Omega_{DE}^{(i)}\approx1$,
we get $G_{eff}\rightarrow\infty$ for both tracker solutions as in \cite{observation-extended-cov-gal-1}.

In the small regime the effective gravitational coupling is approximated
by the relation
\begin{equation}
G_{eff}\approx1-\frac{3(\alpha-2)}{2\left(\Omega_{r}+5\right)}r_{3}^{2}r_{2},
\end{equation}
while in the regime where $r_{3}$ dominates over $r_{1}$ and $r_{2}$
and using $r_{1}=(2-\alpha)/2$, $G_{eff}$ takes the form
\begin{equation}
G_{eff}\approx G_{N}\left[1-\frac{3(\alpha-2)(\Omega_{r}+9)}{2\left(\Omega_{r}+5\right)^{2}}r_{3}^{2}r_{2}\right].
\end{equation}
It follows that in the range $\alpha<-1$, the gravitational interaction
is stronger than that in General Relativity.

\section{Observational constraints\label{sec:Observational-constraints}}

In this section, we place observational bounds on the BG model by
performing a Markov Chain Monte Carlo (MCMC) integration via the Metropolis-Hasting
algorithm. To get tighter constraints we use recent compilation of the redshift space distortion
(RSD) datasets (for our purpose we only consider  the bottom 20 data in Table. \ref{t3}) \cite{perivolaropoulos,perivolaropoulos2}, 
combined with the model-independent observational Hubble (OHD) dataset
obtained through the differential age method \cite{R.Jimenez} shown
in Table. \ref{t4}. The resulting combined Likelihood function reads
\begin{equation}
\mathcal{L}\left(\hat{\theta}\right)=\mathcal{L}_{RSD}\times\mathcal{L}_{OHD}
\end{equation}
and $\hat{\theta}$ is the vector of model parameters over which the
MCMC integration is performed. We consider two cases of the space
parameters for the BG model. In the first case labeled as BG1 model
the space parameter is given by the vector $\hat{\theta}=\left(\alpha,\,r_{1i},\,r_{2i},\,r_{3i},\,\Omega_{m0},\,\sigma_{8}\right)$
and the second case which follows from the dark energy solution found
in \ref{subsec:Dark-energy-solution} is labeled as BG2 model. The
later and is described by the space parameter given by $\hat{\theta}=\left(\alpha,\,r_{2i},\,r_{3i},\,\Omega_{m0},\,\sigma_{8}\right)$,
where the initial condition on $r_{1}$ is fixed as $r_{1}^{(s)}=\left(2-\alpha\right)/2.$
The BG2 model has one parameter less than the BG1 model and it is
expected that it will be less penalized than the BG1 model by Bayesian
selection. Without using any fiducial cosmology to correct the $f\sigma_{8}$
measurements, we confront our findings with the $68$\%, $95$\% and
$99$\% confidence limits of $\textrm{\ensuremath{\Lambda}CDM}$ and
$w$CDM models.

\subsection{Constrained parameter space}

The best fits results of the parameters $r_{1}^{(s)},\:r_{2}^{(s)},\:r_{3}^{s)},\:\alpha,\:\Omega_{m},\:h,\:w$
and $\sigma_{8}$ with $68$\% Confidence Level (CL) limits for the
BG1, BG2, $\textrm{\ensuremath{\Lambda}CDM}$ and $\textrm{wCDM}$
models are summarized in Table. \ref{t2}. The constrained parameters are
compatible with that of $\textrm{\ensuremath{\Lambda}CDM}$ and $w$CDM
models, and are consistent with Planck collaboration data \cite{ade,aghanim2018}.
We note also the similarity between the BG1 and BG2 models. But taking
into account that the BG2 model has one less parameter than the BG1
model, the BG2 it is slightly favored.

\begin{figure}[H]
\begin{center}
\includegraphics[width=10cm,height=6cm]{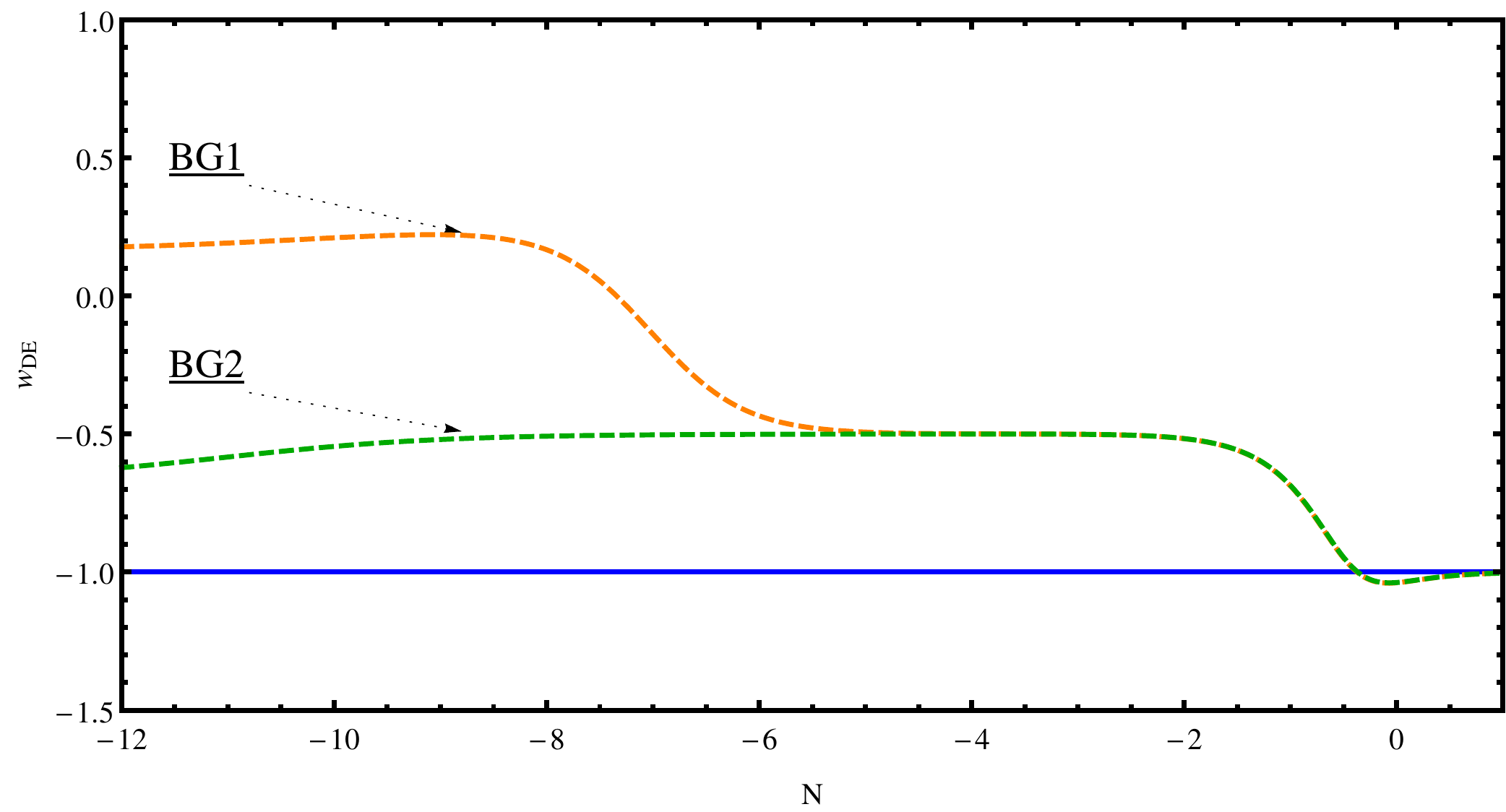}
\caption{\label{Fig.a}Plot of the best fits evolution of the dark energy equation
of state $w_{De}$ for BG1 and BG2 models with respect to $N=\ln a$.
The  best fit parameters used for this plot are given in Table.\ref{t2}.}
\end{center}
\end{figure}

In Fig. \ref{Fig.a} we have shown the best fits evolution of the
dark energy equation of state $w_{DE}$ for BG1 and BG2 models. As
already noted above, $w_{DE}$ follow two different paths in the radiation
epoch until they merge in one path in the matter epoch. The evolution
of $w_{DE}$ in BG2 model is better than that of BG1 model since it
is close to the $\Lambda$CDM model and is in the region $-0.5<w_{DE}<-1.044$
during all cosmological epochs.

\begin{figure}[H]
\begin{center}
\includegraphics[width=10cm,height=6cm]{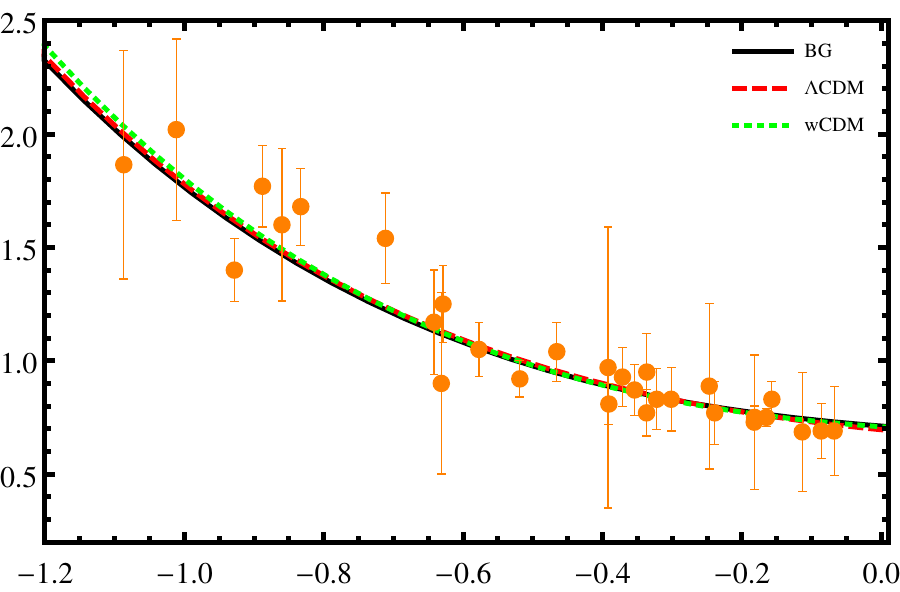}
\caption{\label{Fig.b}Plot of the best fits evolution of the Hubble parameter
for the BG2, $\textrm{\ensuremath{\Lambda}CDM}$ and $w\textrm{CDM}$
models with respect to $N=\ln a$. We have also plotted the Hubble
data with $1\sigma$ errors from the OHD data compilation \cite{R.Jimenez}.}
\end{center}
\end{figure}

In Fig. \ref{Fig.b} we have plotted the evolution of Hubble parameter
for the BG2 model and compared it with the $\textrm{\ensuremath{\Lambda}CDM}$
model and the OHD data. We observe that the curves are indistinguishable
at low redshifts, but they start to differ slightly at high redshifts.

\begin{figure}[H]

\begin{minipage}[b]{0.4\textwidth}
\includegraphics[width=7cm,height=6cm]{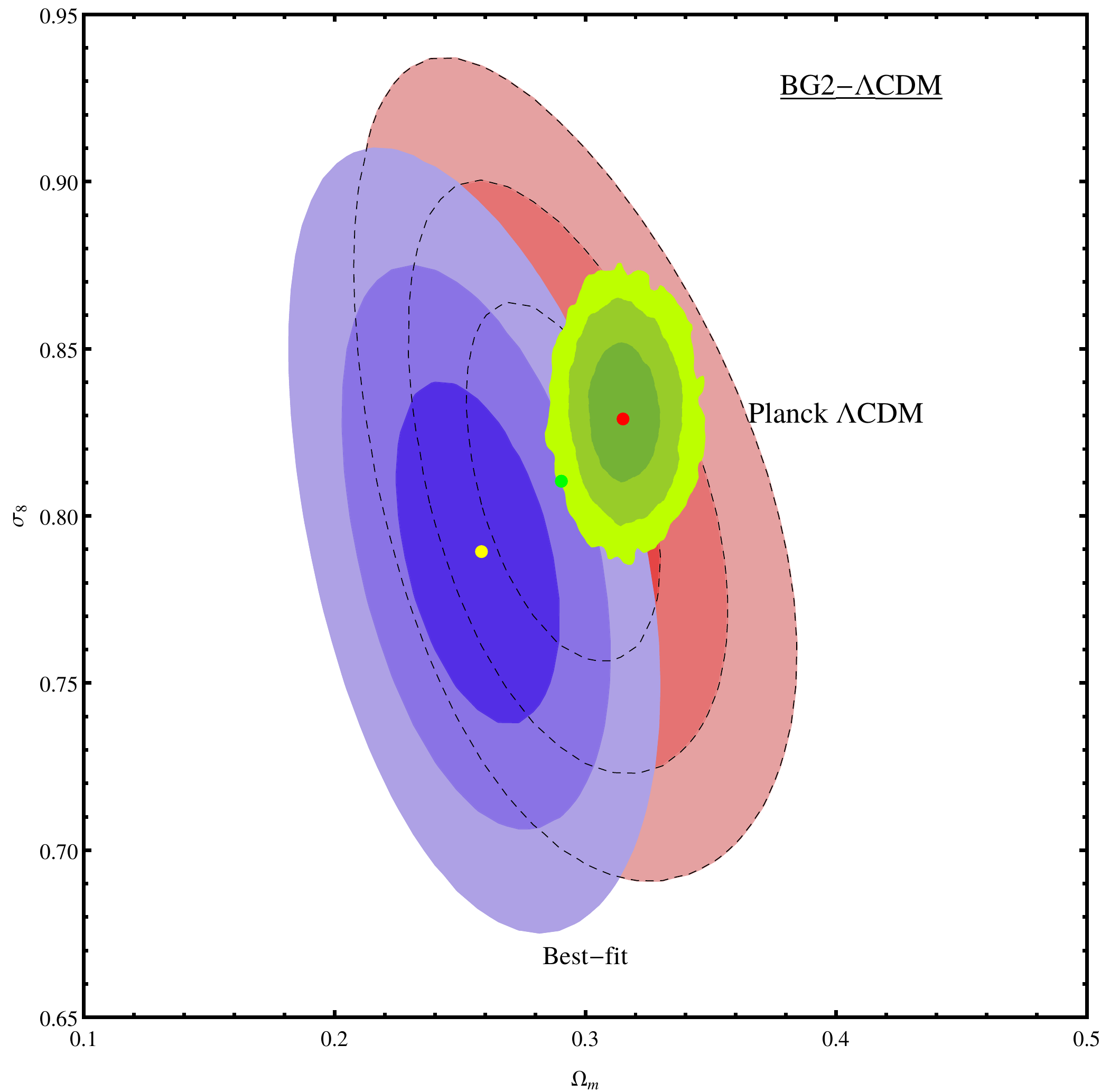}
\end{minipage}
 \hspace*{2cm}
\begin{minipage}[b]{0.4\textwidth}
\includegraphics[width=7cm,height=6cm]{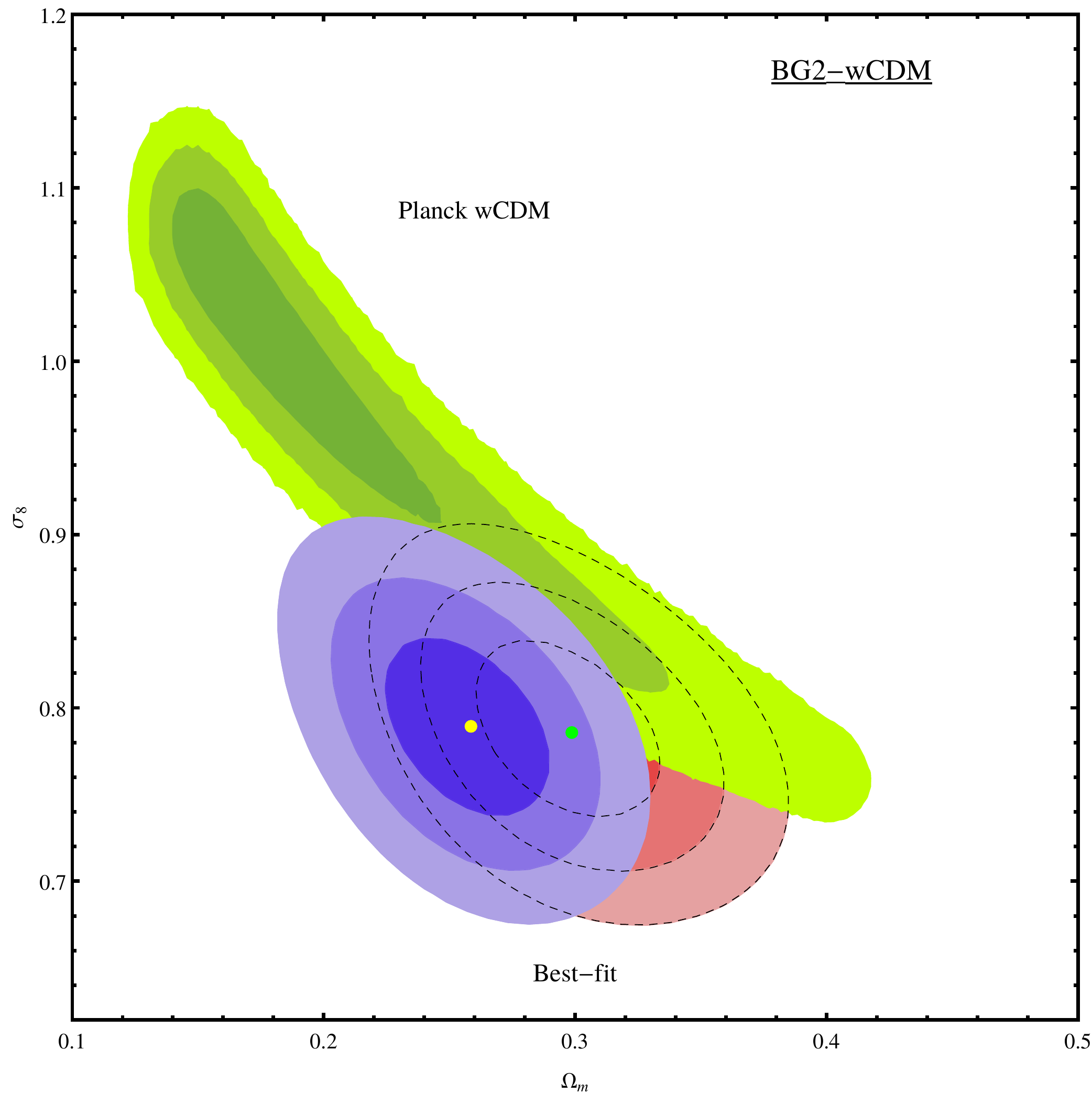}
\end{minipage}

\caption{\label{Fig.c}Probability contours in the $\left(\Omega_{m},\sigma_{8}\right)$
-plane for BG2, $\textrm{\ensuremath{\Lambda}CDM}$ and $w\textrm{CDM}$
models from combined RSD data and OHD data. The filled dark, medium
and light colored contours enclose 68.3, 95.4 and 99.7\% of the probability,
for BG2 model (blue) and $\textrm{CDM}$ models (Red), respectively.
The light Green contours correspond the Planck15/$\textrm{\ensuremath{\Lambda}CDM}$\cite{ade}.}
\end{figure}

We also plotted the observationally allowed regions with $1\sigma,\:2\sigma$
and $3\sigma$ CL limits for parameters $\alpha,\:\Omega_{m},\:h,\:w$
and $\sigma_{8}$ for the BG2, $\textrm{\ensuremath{\Lambda}CDM}$
and $w$CDM models. In Fig. \ref{Fig.c}, the combined recent RSD
and OHD datasets lead the best fit values with $1\sigma$ error for
$\sigma_{8}$ and $\Omega_{m}^{(0)}$ as $(0.7968_{-0.0148}^{+0.0148},\:0.2586_{-0.0277}^{+0.0277})$
for the BG model, and are at $1.5\,\sigma$ from Planck15 values.
The best fit values found for the $\Lambda$CDM, $(0.8104_{-0.0407}^{+0.0407},\:0.2926_{-0.0319}^{+0.0319})$,
are in agreement within $1\,\sigma$ with Planck15 values. Recently,
$\sigma_{8}$ and $\Omega_{m}^{(0)}$ for a $\Lambda$CDM cosmology,
have been constrained by tomographic weak gravitational lensing data
in KiDS-450 survey \cite{Hildebrandt}, and by galaxy clustering and
weak gravitational lensing data from the first year of the Dark Energy
Survey (DES) \cite{abbott} through the relation $S_{8}=\sigma_{8}\sqrt{\Omega_{m}^{(0)}/0.3}$
. In KiDS-450 survey they found $S_{8}=0.745\pm0.039,$ which is at
2.3$\sigma$ from Planck results, while in DES survey the best fit
value, $S_{8}=0.783_{-0.025}^{+0.021}$, is within $1\sigma$ region
of Planck results. In order to adapt the predictions from these datasets
for the BG model we follow \cite{barros}. The value of $S_{8}$ at
a given red-shift for the $\Lambda$CDM cosmology is given by
\begin{flalign}
S_{8(\Lambda)}\left(\bar{z}\right)= & \sigma_{8(\Lambda)}g_{(\Lambda)}\left(\bar{z}\right)\sqrt{\frac{\Omega_{m(\Lambda)}\left(\bar{z}\right)}{0.3}}\nonumber \\
= & S_{8(\Lambda)}g_{(\Lambda)}\left(\bar{z}\right)\sqrt{\frac{\Omega_{m(\Lambda)}\left(\bar{z}\right)}{\Omega_{m(\Lambda)}^{(0)}},}
\end{flalign}
 where $\bar{z}$ is some averaged redshift, $g\left(z\right)=\delta\left(z\right)/\delta\left(0\right)$
of the growth function, and the subscript $\left(\Lambda\right)$
stands for $\Lambda$CDM quantities.

For the BG model we use the rescaled relations \cite{barros}
\begin{flalign}
S_{8}= & S_{8(\Lambda)}\frac{g_{(\Lambda)}\left(\bar{z}\right)}{g\left(\bar{z}\right)}\sqrt{\frac{\Omega_{m(\Lambda)}\left(\bar{z}\right)}{\Omega_{m}\left(z\right)}}\sqrt{\frac{\Omega_{m}^{(0)}}{\Omega_{m(\Lambda)}^{(0)}}},\label{eq:s8}\\
\sigma_{8}= & \sigma_{8(\Lambda)}\frac{g_{(\Lambda)}\left(\bar{z}\right)}{g\left(\bar{z}\right)}\sqrt{\frac{\Omega_{m(\Lambda)}\left(\bar{z}\right)}{\Omega_{m}\left(z\right)}}.\label{eq:sigma8}
\end{flalign}
Considering an average value $\bar{N}=-0.4055\:\left(\bar{z}=0.5\right)$,
we get the following best fit values
\begin{flalign*}
S_{8}=0.734\pm0.081,\: & \sigma_{8}=0.791\pm0.086\quad\textrm{for KiDS-450 survey,}\\
S_{8}=0.771\pm0.078,\: & \sigma_{8}=0.831\pm0.082\quad\textrm{for DES survey.}
\end{flalign*}
In the case of KiDS-450 data, the best fit value for $\sigma_{8}$
is in agreement with the value we have obtained using RSD+H$\left(z\right)$
measurements, and still compatible with Planck values. On the other
the $\sigma_{8}$ value in the context of DES survey does not agree
with the best fit value we found with RSD+H$\left(z\right)$ measurements,
but it is within the $1\sigma$ region of Planck results. We conclude
that the best fit BG2 model is in concordance with both data, but
it is slightly favored if we used instead the DES survey data.

Finally, considering the $w$CDM model we found the best fit values,
$\sigma_{8}=0.7858_{-0.0597}^{+0.0597}$ and $\Omega_{m}^{(0)}=0.2987_{-0.0402}^{+0.0402}$
which are are also close to the $1\sigma$ region of Planck values.
Our best fit values are at $1.5\sigma,\:0.81\sigma$ and $1.1\sigma$
from Planck, $\Lambda$CDM, and $w$CDM best fit values, respectively.
Therefore, we conclude that the BG2 provides a rate of structure clustering
in agreement with current observations.

\begin{figure}[H]

\begin{minipage}[b]{0.4\textwidth}
\includegraphics[width=7cm,height=6cm]{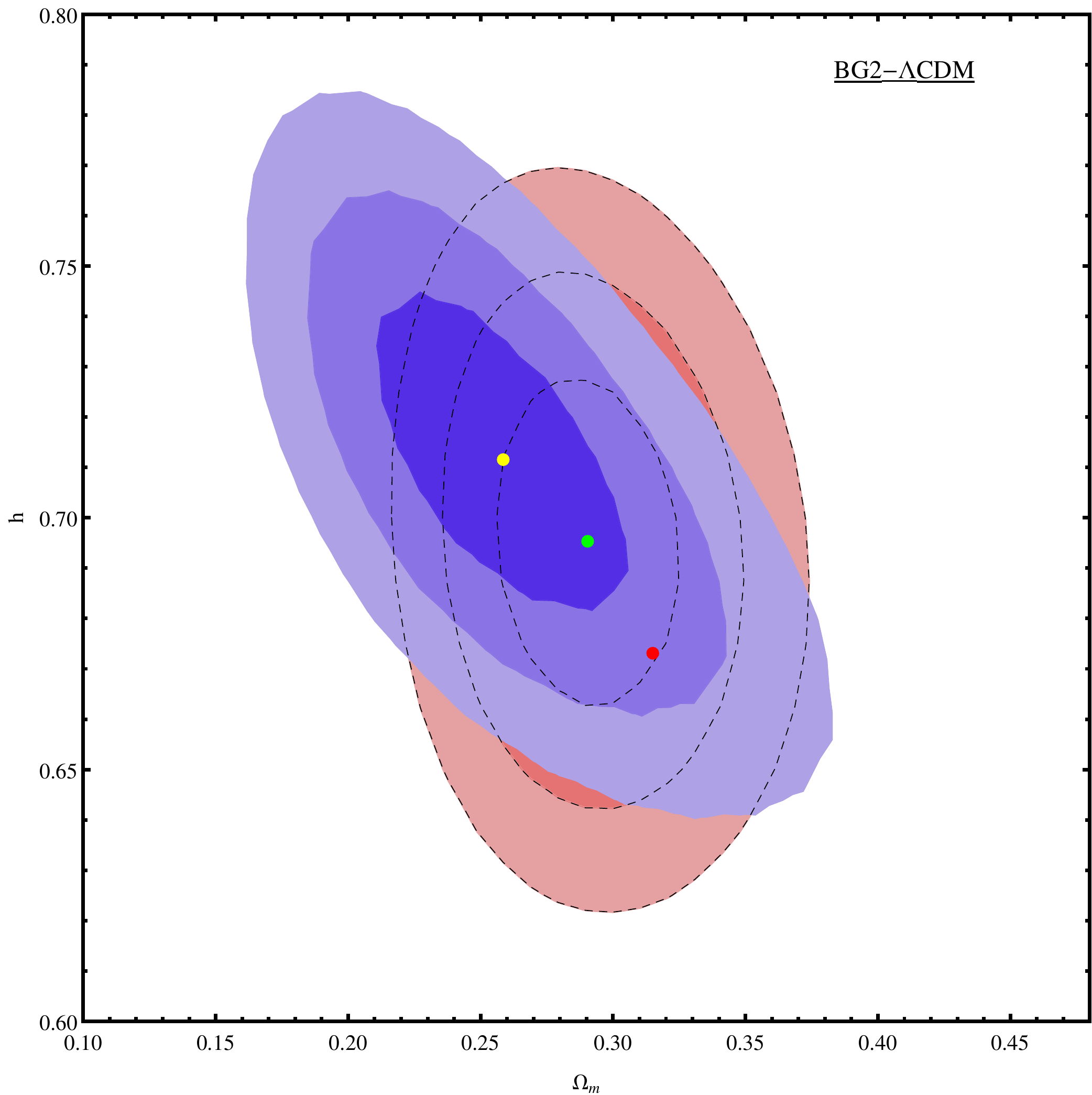}
\end{minipage}
\hspace*{2cm}
\begin{minipage}[b]{0.4\textwidth}
\includegraphics[width=7cm,height=6cm]{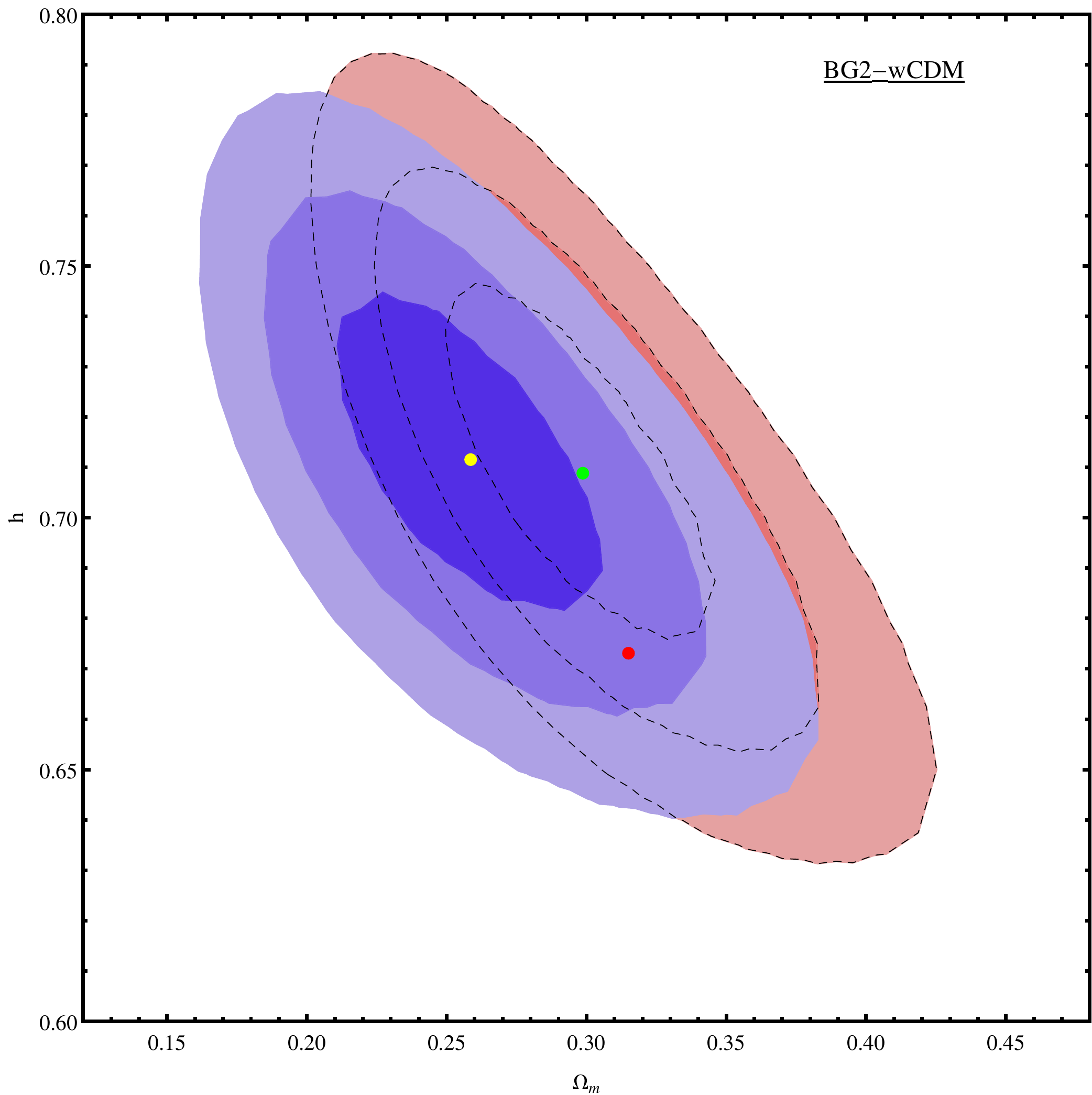}
\end{minipage}

\caption{\label{Fig.d}Probability contours in the $\left(\Omega_{m},h\right)$-planes
for BG2, $\textrm{\ensuremath{\Lambda}CDM}$ and $w\textrm{CDM}$
models from combined RSD data and OHD data. The filled dark, medium
and light colored contours enclose 68.3, 95.4 and 99.7\% of the probability,
for BG2 model (blue) and $\textrm{CDM}$ models (Red), respectively.
The light Green contours correspond the Planck15/$\textrm{\ensuremath{\Lambda}CDM}$\cite{ade}.}
\end{figure}

\begin{figure}[H]

\begin{minipage}[b]{0.4\textwidth}
\includegraphics[width=7cm,height=6cm]{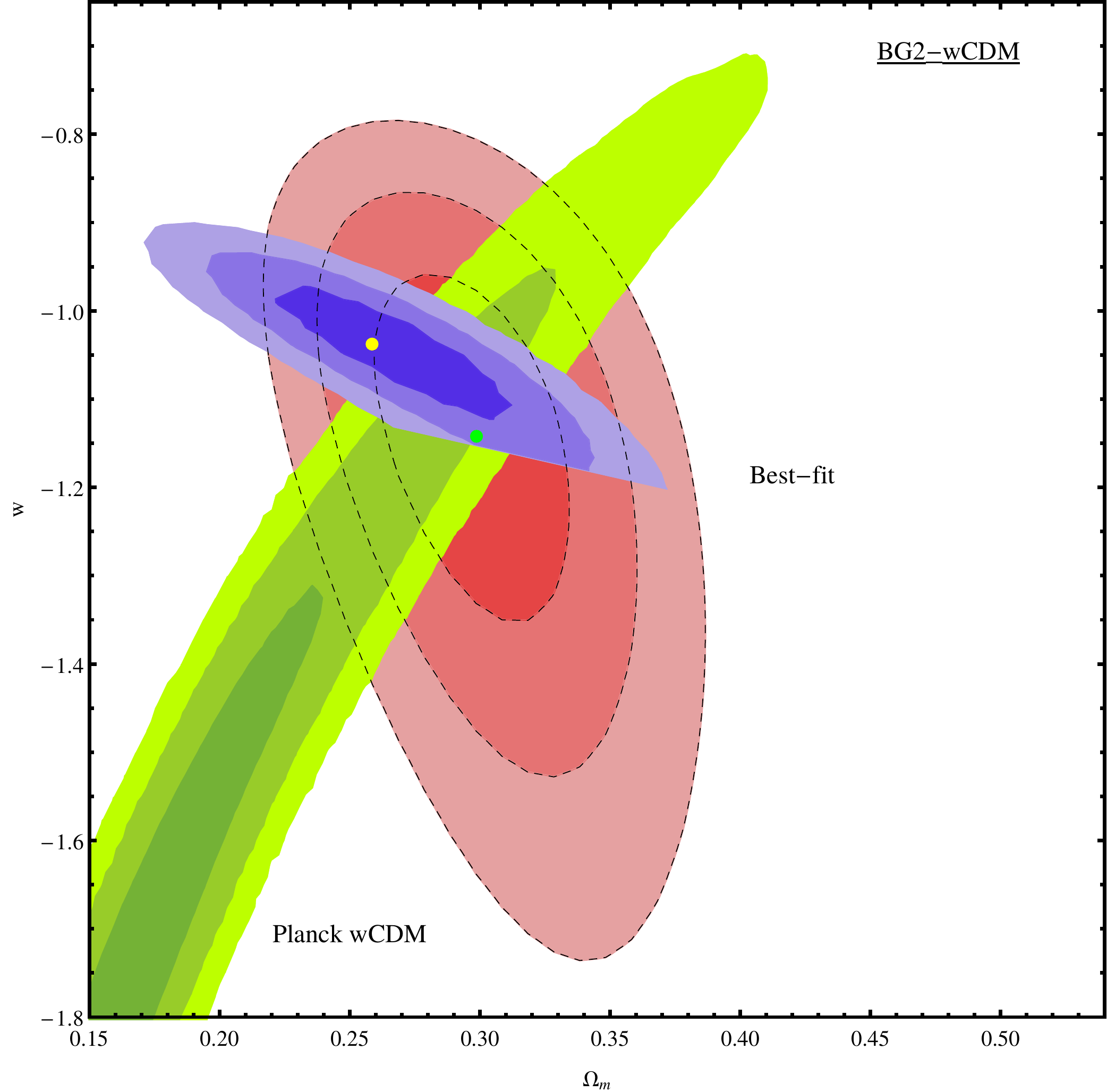}
\end{minipage}
 \hspace*{2cm}
\begin{minipage}[b]{0.4\textwidth}
\includegraphics[width=7cm,height=6cm]{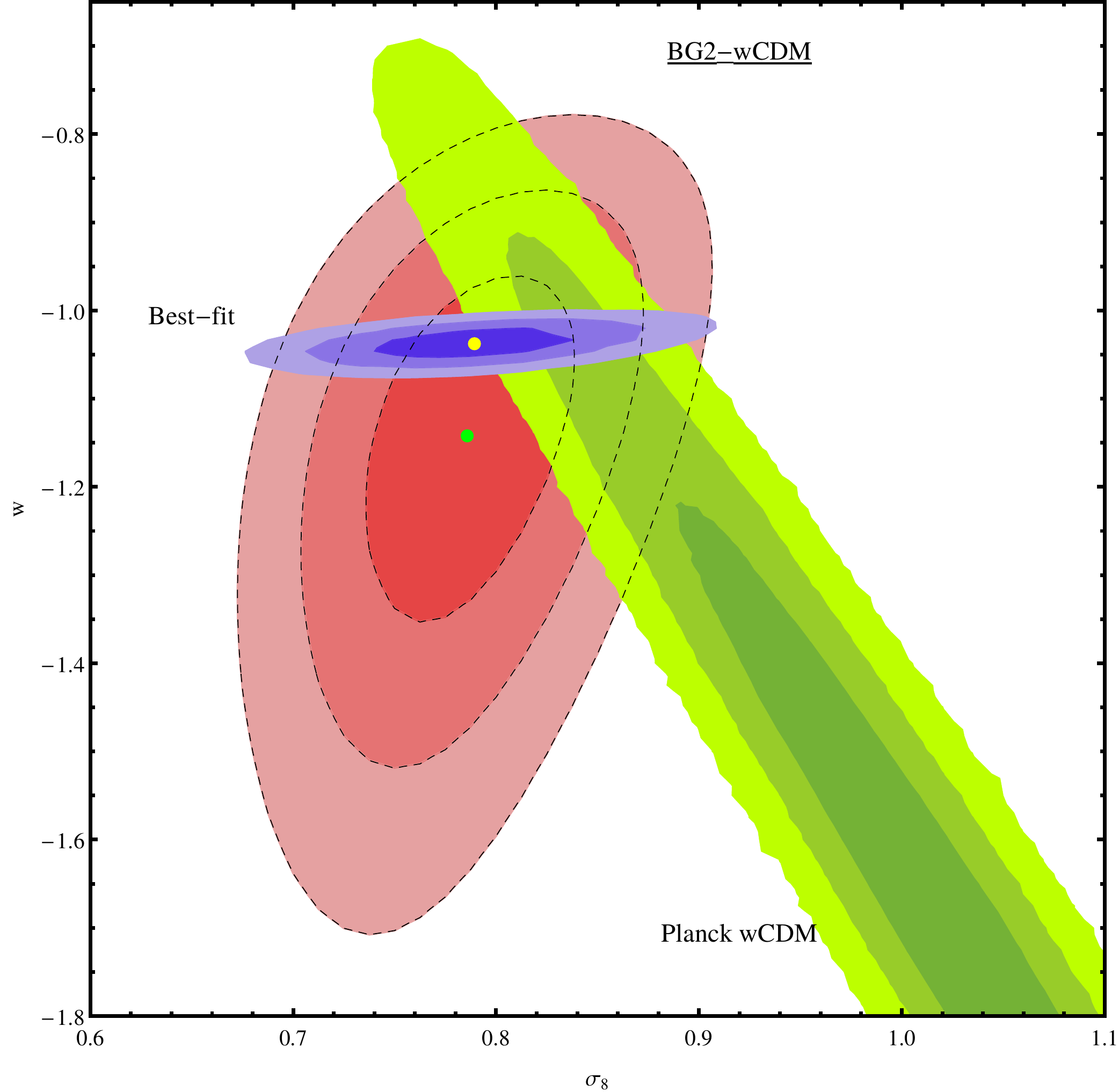}
\end{minipage}

\caption{\label{Fig.e}Probability contours in the $\left(\Omega_{m},w\right)$
and $\left(\sigma_{8},w\right)$-planes for BG2 and $\textrm{wCDM}$
models from combined RSD data and OHD data. The filled dark, medium
and light colored contours enclose 68.3, 95.4 and 99.7\% of the probability,
for BG2 model (blue) and $\textrm{wCDM}$ model (Red), respectively.
The light Green contours correspond the Planck15/$\textrm{wCDM}$\cite{ade}.}
\end{figure}

In Fig. \ref{Fig.d}, the combined data RSD+H$\left(z\right)$ lead
to higher values of $h$ at $1\sigma$ $\left(h=0.7116_{-0.0288}^{+0.0288}\right)$,
in the range of the values found in \cite{BVRiess,AdRiess,HuangRiess}. This value is very close to the recent local measurement of Hubble constant, $H_0=69.8 \pm 0.8$ \cite{WLFreedmann2}, and   
therefore eases the persistent tension on the Hubble constant.
In Fig. \ref{Fig.e}, we show the
data constraints on the today dark energy equation of state $w_{0}$.
The best fit value for the BG2 model, $w_{0}=-1.0377_{-0.068}^{+0.068}$
is very close to $-1$, and that the BG2 model is more constrained
by the data than the $w$CDM model regarding the constraints from
Planck15/$w$CDM.

\begin{figure}[H]

\begin{minipage}[b]{0.4\textwidth}
\includegraphics[width=7cm,height=6cm]{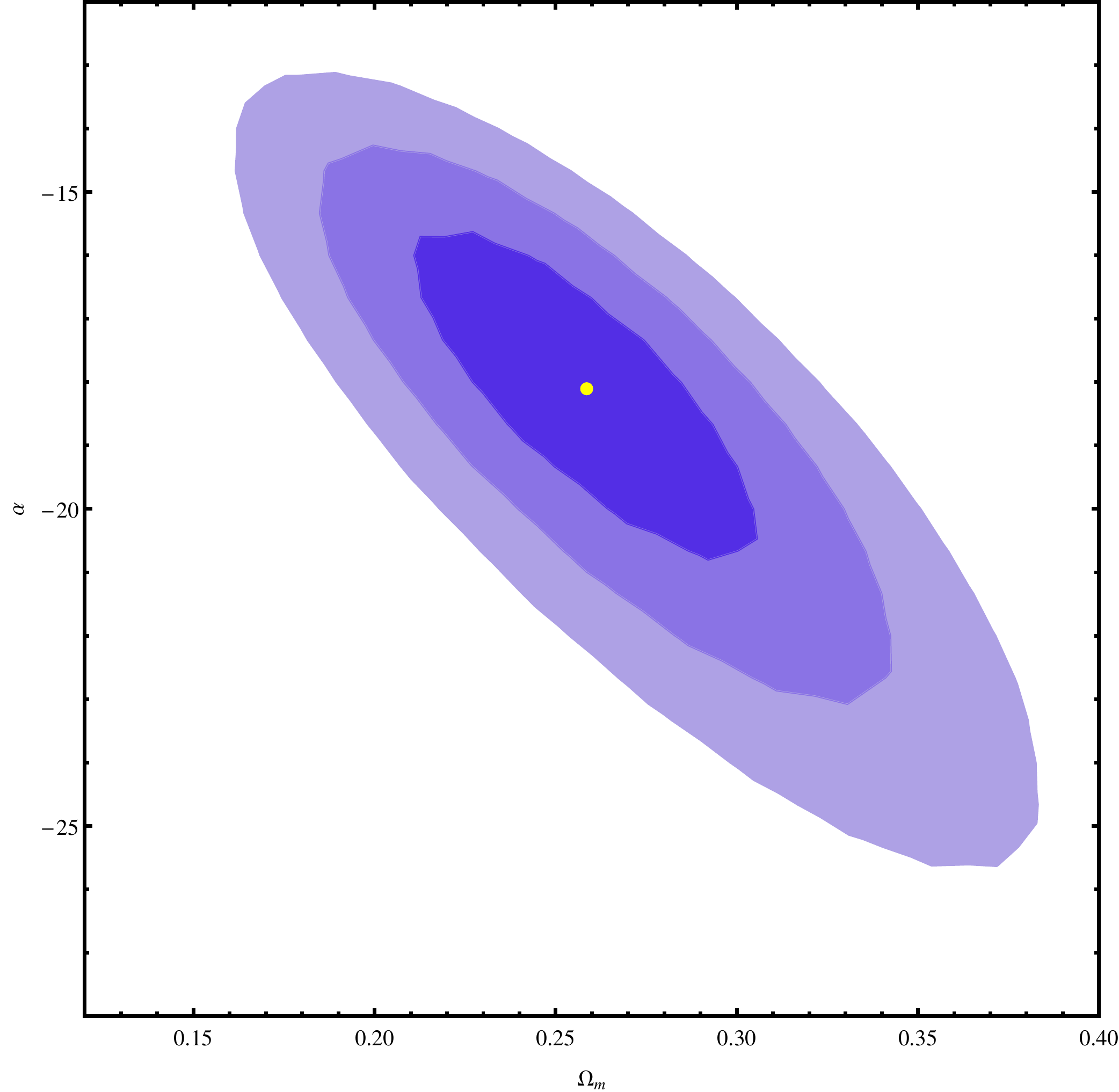}
\end{minipage}
 \hspace*{2cm}
\begin{minipage}[b]{0.4\textwidth}
\includegraphics[width=7cm,height=6cm]{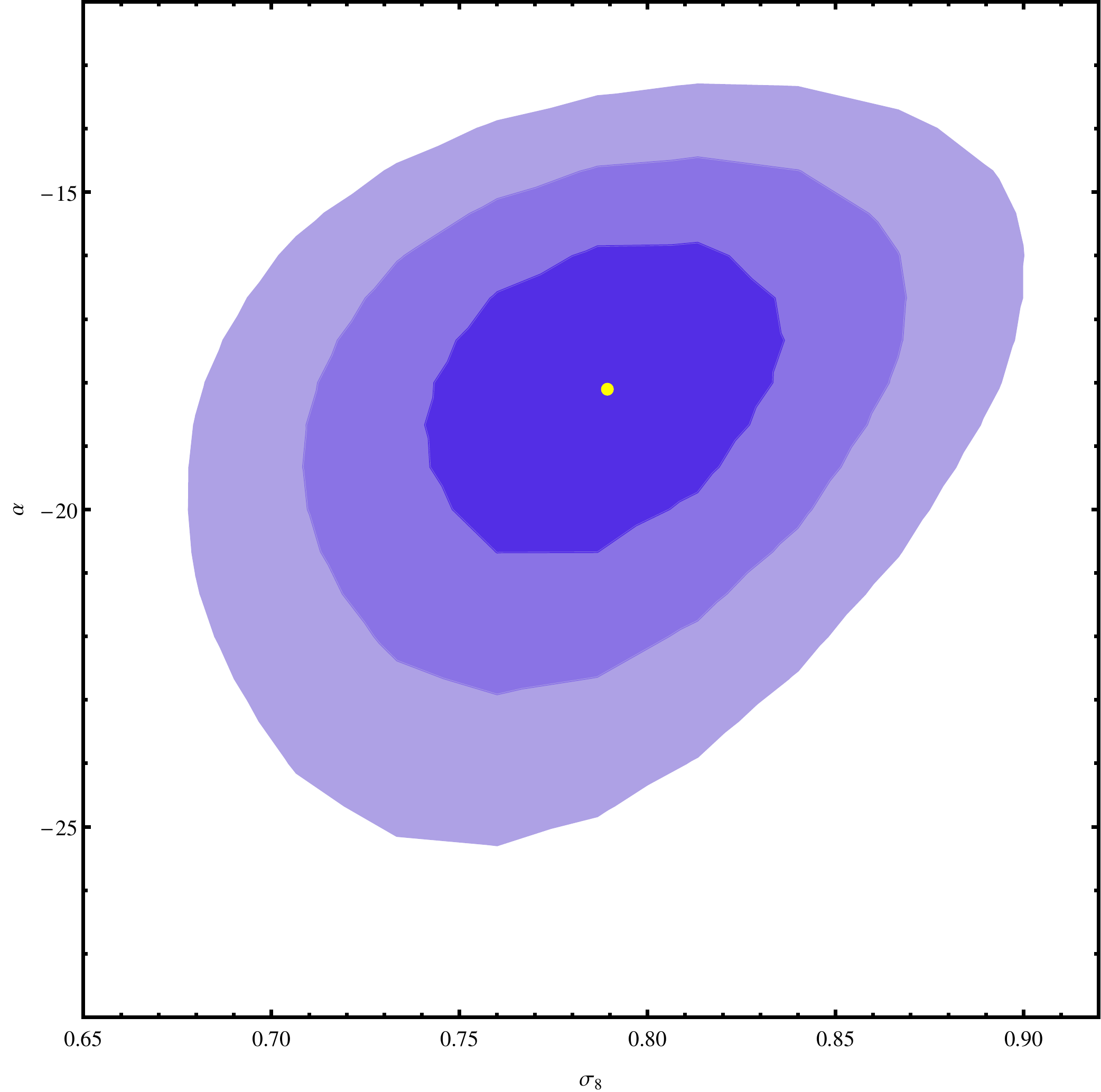}
\end{minipage}

\caption{\label{Fig.f}Probability contours in the $\left(\Omega_{m},\alpha\right)$,
$\left(\sigma_{8},\alpha\right)$-planes for BG2 model from combined
RSD data and OHD data. The filled dark, medium and light blue contours
enclose 68.3, 95.4 and 99.7\% of the probability, respectively.}
\end{figure}

In Fig. \ref{Fig.f}, the probability contours in the $\left(\Omega_{m},\alpha\right)$
and $\left(\sigma_{8},\alpha\right)$-planes show that the coupling
constant is constrained by the data to $\alpha=-18.1045_{-3.366}^{+3.366}$
.

\begin{figure}[H]
\begin{center}
\includegraphics[width=15cm,height=6cm]{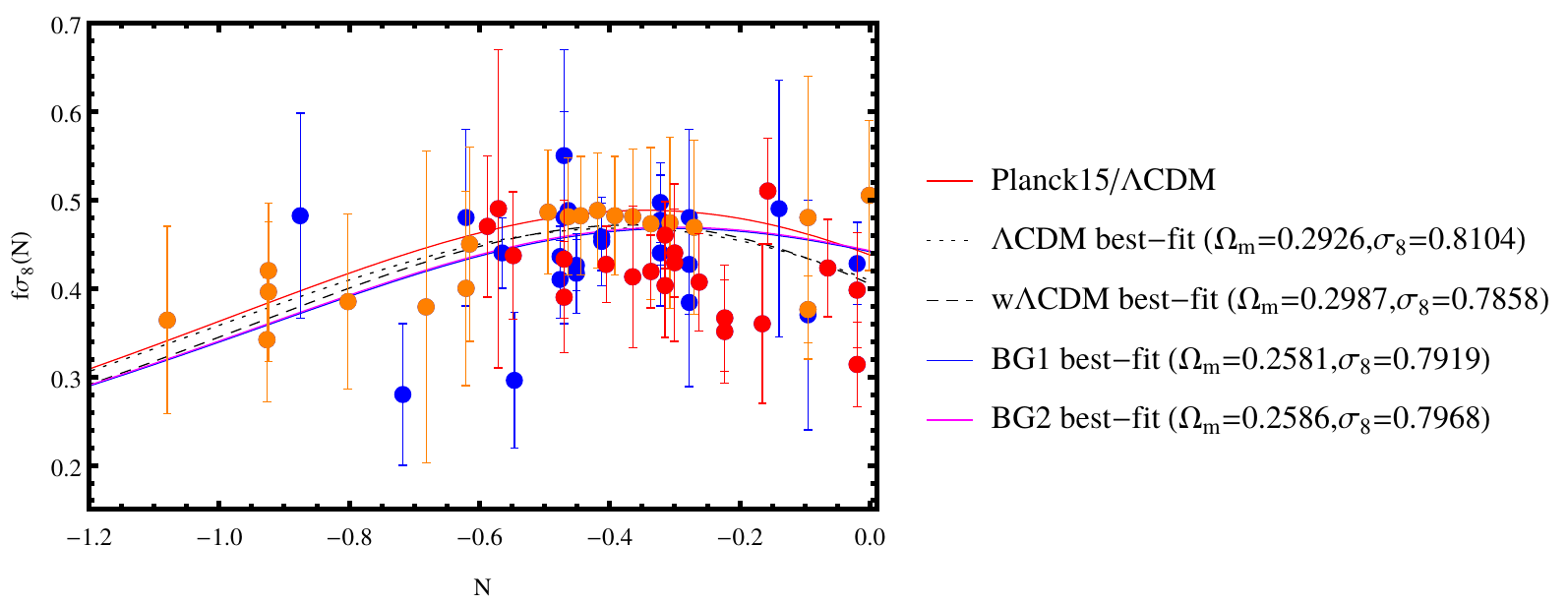}
\caption{\label{Fig.g}Plot of $f\sigma_{8}\left(N\right)$ as a function of
the number of e-folds for the 20 latest growth rate data set (Orange).
We also show the earliest data points (Red) and the remaining published
data points (Blue) taken from Table III. The red curve corresponds
to Planck15/$\Lambda\textrm{CDM best fits.}$The indistinguishable
blue and cyan solid lines corresponds to the best fits BG1 and BG2
models respectively, whereas the dotted and dashed black lines corresponds
to the best fits $\Lambda\textrm{CDM }$and $w\textrm{CDM}$ models,
respectively.}
\end{center}
\end{figure}

Fig. \ref{Fig.g} shows the best fit behavior of $f\sigma_{8}\left(N\right)$
for the BG1 and BG2 models. The two models are indistinguishable,
and we notice that the strenght of fluctuations is stronger than that
of the $\Lambda$CDM and $w$CDM models starting from redshift $z\approx0.41,\:0.35$
to the prsent epoch, respectively. This means that the structures
cluster faster in the BG model than in the $\Lambda$CDM and $w$CDM
models in this recent past epoch, and this effect is due to increasing
behavior of the effective gravitational constant. Compared to Planck15
data, the strength of the fluctuations in the BG models becomes stronger
only starting from $z\apprge$  $0.004.$ Even we have only considered
in our study the RSD data published recently, it is this particular
behavior of matter fluctuations in the BG model which make our results
are still consistent with the full growth data. Finally, we provide
a parametrization for $f\sigma_{8}\left(z\right)$ in the BG2 model
assuming the $\Lambda$CDM background in the form $f\sigma_{8}\left(z\right)=\rho\sigma_{8}\Omega_{m}\left(z\right)^{\gamma}/\left(1+z\right)^{\beta}$.
Using Planck15 data we obtain an excellent fit to the numerical solution
of Eqs.(\ref{eq:delta}) and (\ref{Geff-1}) with the best fit parameters
$\rho\approx1.16,\:\gamma\approx0.6$ and $\beta\approx0.93.$ These
values are very close to that obtained in modified gravity theory
parametrization of $G_{eff}\left(z\right)$ \cite{perivolaropoulos,perivolaropoulos2}.

\begin{table}
\begin{centering}
\begin{tabular*}{16cm}{@{\extracolsep{\fill}}c|c|c|c|c}
\hline
\textbf{\scriptsize{Parameter}} & {\scriptsize{}$\begin{array}{c}
\underline{\textrm{BG1}}\\
\chi_{min}^{2}=21.8994
\end{array}$} & {\scriptsize{}$\begin{array}{c}
\underline{\textrm{BG2}}\\
\chi_{min}^{2}=21.8795
\end{array}$} & {\scriptsize{}$\begin{array}{c}
\underline{\textrm{\ensuremath{\Lambda}CDM}}\\
\chi_{min}^{2}=19.8405
\end{array}$} & {\scriptsize{}$\begin{array}{c}
\underline{\textrm{\ensuremath{w}CDM}}\\
\chi_{min}^{2}=19.83
\end{array}$}\tabularnewline
\hline
{\scriptsize{}$\alpha$} & {\scriptsize{}$-18.1\pm3.4658$} & {\scriptsize{}$-18.1045\pm3.3660$} & {\scriptsize{}$-$} & {\scriptsize{}$-$}\tabularnewline
{\scriptsize{}$r_{1i}\left(.10^{-14}\right)$} & {\scriptsize{}$26.8058\pm5.3154$} & {\scriptsize{}$-$} & {\scriptsize{}$-$} & {\scriptsize{}$-$}\tabularnewline
{\scriptsize{}$r_{2i}\left(.10^{-22}\right)$} & {\scriptsize{}$6.45\pm2.42323$} & {\scriptsize{}$\left(6.734\pm2.4047\right)\times10^{-40}$} & {\scriptsize{}$-$} & {\scriptsize{}$-$}\tabularnewline
{\scriptsize{}$r_{3i}\left(.10^{6}\right)$} & {\scriptsize{}$8.2042\pm2.7431$} & {\scriptsize{}$\left(2.5\pm0.9306\right)\times10^{14}$} & {\scriptsize{}$-$} & {\scriptsize{}$-$}\tabularnewline
{\scriptsize{}$\Omega_{m}^{(0)}h^{2}$} & {\scriptsize{}$0.1305\pm0.01$} & {\scriptsize{}$0.1309\pm0.009$} & {\scriptsize{}$0.1406\pm0.0136$} & {\scriptsize{}$0.15\pm0.0219$}\tabularnewline
{\scriptsize{}$\sigma_{8}$} & {\scriptsize{}$0.7919\pm0.0177$} & {\scriptsize{}$0.7968\pm0.0148$} & {\scriptsize{}$0.8104\pm0.0407$} & {\scriptsize{}$0.7858\pm0.0597$}\tabularnewline
{\scriptsize{}$h$} & \multirow{1}{*}{{\scriptsize{}$0.7111\pm0.0298$}} & {\scriptsize{}$0.7116\pm0.0288$} & {\scriptsize{}$0.6953\pm0.0246$} & {\scriptsize{}$0.7088\pm0.0519$}\tabularnewline
{\scriptsize{}$w_{0}$} & {\scriptsize{}$-1.0385\pm0.0756$} & {\scriptsize{}$-1.0377\pm0.0682$} & {\scriptsize{}$-1$} & {\scriptsize{}$-1.1421\pm0.3465$}\tabularnewline
{\scriptsize{}$\Omega_{m}^{(0)}$} & {\scriptsize{}$0.2581\pm0.028$} & {\scriptsize{}$0.2586\pm0.0277$} & {\scriptsize{}$0.2926\pm0.0319$} & {\scriptsize{}$0.2987\pm0.0402$}\tabularnewline
\hline
{\scriptsize{}$\ln B_{ij}$/$\Lambda$CDM} & {\scriptsize{}$-0.5$} & {\scriptsize{}$-0.25$} & {\scriptsize{}$-$} & {\scriptsize{}$-0.33$}\tabularnewline
\hline
{\scriptsize{}$\ln B_{ij}$/$w$CDM} & {\scriptsize{}$-0.16$} & {\scriptsize{}$0.08$} & {\scriptsize{}$0.33$} & {\scriptsize{}$-$}\tabularnewline
\hline
\end{tabular*}
\par\end{centering}
\caption{\label{t2}1$\sigma$ Parameter confidence level for the BG, $\Lambda$CDM
and $w$CDM models from late RSD dataset combined
with OHD dataset. The parameters $h$ and $w$ are derived parameters
for the BG models, while $\Omega_{m0}$ is a derived parameter for
all the models. The negative values of $\textrm{ln}B_{ij}$ imply
that $\Lambda$CDM and $w$CDM models are preferred over the BG1
and BG2 models.}
\end{table}

\subsection{Bayesian analysis}

In order to see whether the BG models are favored over the $\textrm{\ensuremath{\Lambda}CDM}$
and $w\textrm{CDM}$ models, we use Bayesian analysis where the central
quantity is the posterior probability of the parameter space $\theta$
subjected to observational data and prior information. Given two models
$\mathcal{M}_{A}$ and $\mathcal{M}_{B}$ the posterior beliefs on
the two models is given by
\begin{equation}
\frac{\textrm{Pr}\left(\mathcal{M}_{A}\mid\mathcal{Y}\right)}{\textrm{Pr}\left(\mathcal{M}_{B}\mid\mathcal{Y}\right)}=\frac{\pi\left(\mathcal{M}_{A}\right)}{\pi\left(\mathcal{M}_{B}\right)}\frac{\textrm{Pr}\left(\mathcal{Y}\mid\mathcal{M}_{A}\right)}{\textrm{Pr}\left(\mathcal{Y}\mid\mathcal{M}_{B}\right)}\label{eq:Bayes1}
\end{equation}
where $\pi\left(\mathcal{M}_{i}\right)$ is the prior belief on the
model $\mathcal{M}_{i}$. The updating term on the RHS of Eq.(\ref{eq:Bayes1}),
$B_{AB}=\textrm{Pr}\left(\mathcal{Y}\mid\mathcal{M}_{A}\right)/\textrm{Pr}\left(\mathcal{Y}\mid\mathcal{M}_{B}\right)$,
is Bayes factor of the model $\mathcal{M}_{A}$ relative to the reference
model $\mathcal{M}_{B}$, and is a key quantity in Bayesian hypothesis
testing \cite{Kass_Raftery}. The Bayes factor provides a direct measure
of the weight of evidence provided by data for the reference model.
A quantification of this measure is provided by the classification
proposed by Kass and Raftery \cite{Kass_Raftery}, where $1<B_{ij}<3$
is interpreted as weak evidence; $3<B_{ij}<20$ as positive evidence;
$20<B_{ij}<150$ as strong evidence; and $B_{ij}>150$ as very strong
evidence in favor of the reference model.

We have calculated the Bayes factor taking the $\Lambda$CDM and $w$CDM
models as the reference models. In the first case the datasets exhibit
the preference of the $\Lambda$CDM with only three free parameters
over the BG1, BG2 and $w$CDM models. On the other hand the BG2 model
with one extra parameter than in $w$CDM model is slightly preferred
with a score $0.08$ of Bayes factor.

\section{Conclusions\label{sec:Conclusion}}

In this paper, we studied the background cosmology and the evolution
of matter density perturbations in a covariant multi-Galileons field
model. In particular, we considered the cubic bi-Galileon (BG) model
with two scalar fields $\varphi^{1}$ and $\varphi^{2}$ and constant
coupling functions. We showed that the BG model exhibits a rich dark
energy structure compared to the single Galileon (SG) model.
The phase space analysis of the dynamical equations of the background
cosmology allowed us to identify a set of fixed points and their properties
in each cosmological epoch, and that the supplementary dynamical variable
$r_{3}=\dot{\varphi}^{2}/\dot{\varphi}^{1}$, compared to the SG field
model, plays a crucial role in the cosmological dynamics of the BG
model. We found that the BG model possess two tracker solutions, one
is the usual tracker solution of the SG field model and the other
is considered as the signature of the second field. We investigate
the properties of the BG model by considering two cases, the first
one, labeled BG1, consists in analyzing the full set of equations
of the dynamical system, and the second case, labeled BG2, is based
on the dark energy solution found in Sec. \ref{subsec:Dark-energy-solution}
valid in the regime where $r_{3}^{(s)}>>\left\{ 1,\:r_{1}^{(s)}\right\} $
and $r_{2}^{(s)}\ll1$ to guaranty $\Omega_{DE}\ll1$ radiation and
matter epochs. We show that the cosmological evolution prefer the
path of the second tracker at earlier or later times depending on
the initial conditions. However, these tracker solutions are disfavored
by observational data due to the bad behavior of the dark energy equation
of state in matter epoch $\left(w_{DE}=-2\right)$, exactly like the
cubic SG model. However, in the regime of initial conditions $r_{3}^{(s)}>>\left\{ 1,\:r_{1}^{(s)}\right\} $
and $r_{2}^{(s)}\ll1$, we found that the approach to the tracker
is prevented in BG1 and BG2 models, respectively. In fact, we showed
that the dark energy solution in this regime with the initial condition
$r_{1}^{(s)}=(2-\alpha)/2$, gives the best sequence of the evolution
of the dark energy equation of state, $w_{DE}=-2/3\:(\textrm{radiation era})\rightarrow-1/2\:(\textrm{matter era})\rightarrow-1\:(\textrm{dS era})$.
We studied also the growth rate of matter perturbation in the quasi-static
approximation on sub-horizon scales, and obtained the evolution equation
of the matter density contrast with an effective gravitational coupling
showing that gravity is stronger than that in General Relativity.
Using the combination of the latest RSD data and the OHD data compilations,
we put observational constraints on free parameters in the BG model
by running the MCMC simulation using Hasting-Metropolis algorithm
assuming flat priors for the fitting parameters. We considered the
two models BG1 and BG2 with six and five parameters, respectively.
The results obtained from the likelihood analysis shows that both
the BG models give practically indistinguishable best fits parameters
$\left(\alpha,\:\Omega_{m},\:w,\:h,\:\sigma_{8}\right)$ where $\alpha$
is the coupling function, and are compatible with the Planck15 (TT+lowE)
observations data. Indeed the best fits value of $\sigma_{8}$ for
the BG2 model, $\sigma_{8}=0.7968\pm0.0148$, is very close to Planck15
$1\sigma$ uncertainties. More interestingly, we have found that the
best fits values of the Hubble constant, $h=0.7116\pm0.0288$, can
ease the persistent tension between CMB \cite{aghanim2018} and Cepheid
distance scale measurements at low redshifts \cite{AdRiess}. Finally,
according to model selection using the Bayes factor, we found that
the BG2 model is disfavored compared to $\Lambda$CDM model but slightly
preferred over the $w$CDM model . In future works, it will be of interest
to place stronger constraints on the initial conditions $r_{i}^{(s)}$
and cosmological parameters using other observational data such Type
Ia Supernovae (SnIa), Baryon Acoustic Oscillation (BAO), Weak Lensing
(WL) and Cosmic Microwave Background (CMB) measurements.

\section*{Acknowledgments}

The research of K. N was supported by the The Algerian Ministry of Higher Education
and Scientific Research grant no. ''CNEPRU-D01720140008''.

\newpage{}

\subsection*{Appendix: Tables of data}

\begin{table}[ht]
\renewcommand{\arraystretch}{0.45}
\centering{}\caption{\label{t3}Redshift Space Distorsion data compilation \cite{perivolaropoulos,perivolaropoulos2}.}
{\scriptsize{}}%
\begin{tabular}{cccc>{\centering}p{2cm}}
\hline
{\scriptsize{}Index} & {\scriptsize{}Data set} & {\scriptsize{}$z$ } & {\scriptsize{}$\mathit{f\sigma_{8}}(z)$} & {\scriptsize{}References}\tabularnewline
\hline
{\scriptsize{}1} & {\scriptsize{}SDSS-LRG} & {\scriptsize{}0.35} & {\scriptsize{}0.440 \textpm{} 0.050} & {\scriptsize{}\cite{p75}}\tabularnewline
{\scriptsize{}2} & {\scriptsize{}VVDS} & {\scriptsize{}0.77} & {\scriptsize{}0.490 \textpm{} 0.18} & {\scriptsize{}\cite{p75}}\tabularnewline
{\scriptsize{}3} & {\scriptsize{}2dFGRS} & {\scriptsize{}0.17} & {\scriptsize{}0.510 \textpm{} 0.060} & {\scriptsize{}\cite{p75}}\tabularnewline
{\scriptsize{}4} & {\scriptsize{}2MRS} & {\scriptsize{}0.02} & {\scriptsize{}0.314 \textpm{} 0.048 } & {\scriptsize{}\cite{p77,p78}}\tabularnewline
{\scriptsize{}5} & {\scriptsize{}SnIa+IRAS} & {\scriptsize{}0.02} & {\scriptsize{}0.398 \textpm{} 0.065 } & {\scriptsize{}\cite{p78,p79}}\tabularnewline
{\scriptsize{}6} & {\scriptsize{}SDSS-LRG-200} & {\scriptsize{}0.25} & {\scriptsize{}0.3512 \textpm{} 0.0583} & {\scriptsize{}\cite{p80}}\tabularnewline
{\scriptsize{}7} & {\scriptsize{}SDSS-LRG-200} & {\scriptsize{}0.37} & {\scriptsize{}0.4602 \textpm{} 0.0378} & {\scriptsize{}\cite{p80}}\tabularnewline
{\scriptsize{}8} & {\scriptsize{}SDSS-LRG-60 } & {\scriptsize{}0.25} & {\scriptsize{}0.3665 \textpm{} 0.0601} & {\scriptsize{}\cite{p80}}\tabularnewline
{\scriptsize{}9} & {\scriptsize{}SDSS-LRG-60 } & {\scriptsize{}0.37} & {\scriptsize{}0.4031 \textpm{} 0.0586} & {\scriptsize{}\cite{p80}}\tabularnewline
{\scriptsize{}10} & \multicolumn{1}{c}{{\scriptsize{}WiggleZ}} & {\scriptsize{}0.44} & {\scriptsize{}0.413 \textpm{} 0.080} & {\scriptsize{}\cite{p46}}\tabularnewline
{\scriptsize{}11} & {\scriptsize{}WiggleZ} & {\scriptsize{}0.60} & {\scriptsize{}0.390 \textpm{} 0.063} & {\scriptsize{}\cite{p46}}\tabularnewline
{\scriptsize{}12} & {\scriptsize{}WiggleZ} & {\scriptsize{}0.73} & {\scriptsize{}0.437 \textpm{} 0.072} & {\scriptsize{}\cite{p46}}\tabularnewline
{\scriptsize{}13} & {\scriptsize{}6dFGS} & {\scriptsize{}0.067} & {\scriptsize{}0.423 \textpm{} 0.055} & {\scriptsize{}\cite{p81}}\tabularnewline
{\scriptsize{}14} & {\scriptsize{}SDSS-BOSS} & {\scriptsize{}0.30} & {\scriptsize{}0.407 \textpm{} 0.055} & {\scriptsize{}\cite{p82}}\tabularnewline
{\scriptsize{}15} & {\scriptsize{}SDSS-BOSS} & {\scriptsize{}0.40} & {\scriptsize{}0.419 \textpm{} 0.041} & {\scriptsize{}\cite{p82}}\tabularnewline
{\scriptsize{}16} & {\scriptsize{}SDSS-BOSS} & {\scriptsize{}0.50} & {\scriptsize{}0.427 \textpm{} 0.043} & {\scriptsize{}\cite{p82}}\tabularnewline
{\scriptsize{}17} & {\scriptsize{}SDSS-BOSS} & {\scriptsize{}0.60} & {\scriptsize{}0.433 \textpm{} 0.067} & {\scriptsize{}\cite{p82}}\tabularnewline
{\scriptsize{}18} & {\scriptsize{}Vipers} & {\scriptsize{}0.80} & {\scriptsize{}0.470 \textpm{} 0.080} & {\scriptsize{}\cite{p83}}\tabularnewline
{\scriptsize{}19} & {\scriptsize{}SDSS-DR7-LRG} & {\scriptsize{}0.35} & {\scriptsize{}0.429 \textpm{} 0.089} & {\scriptsize{}\cite{p84}}\tabularnewline
{\scriptsize{}20} & {\scriptsize{}GAMA} & {\scriptsize{}0.18} & {\scriptsize{}0.360 \textpm{} 0.090} & {\scriptsize{}\cite{p86}}\tabularnewline
{\scriptsize{}21} & {\scriptsize{}GAMA} & {\scriptsize{}0.38} & {\scriptsize{}0.440 \textpm{} 0.060} & {\scriptsize{}\cite{p86}}\tabularnewline
{\scriptsize{}22} & {\scriptsize{}BOSS-LOWZ} & {\scriptsize{}0.32} & {\scriptsize{}0.384 \textpm{} 0.095} & {\scriptsize{}\cite{p87}}\tabularnewline
{\scriptsize{}23} & {\scriptsize{}SDSS DR10 and DR11 } & {\scriptsize{}0.32} & {\scriptsize{}0.48 \textpm{} 0.10} & {\scriptsize{}\cite{p87}}\tabularnewline
{\scriptsize{}24} & {\scriptsize{}SDSS DR10 and DR11 } & {\scriptsize{}0.57} & {\scriptsize{}0.417 \textpm{} 0.045} & {\scriptsize{}\cite{p87}}\tabularnewline
{\scriptsize{}25} & {\scriptsize{}SDSS-MGS} & {\scriptsize{}0.15} & {\scriptsize{}0.490 \textpm{} 0.145} & {\scriptsize{}\cite{p89}}\tabularnewline
{\scriptsize{}26} & {\scriptsize{}SDSS-veloc} & {\scriptsize{}0.10} & {\scriptsize{}~0.370 \textpm{} 0.130~} & {\scriptsize{}\cite{p90}}\tabularnewline
{\scriptsize{}27} & {\scriptsize{}FastSound} & {\scriptsize{}1.40} & {\scriptsize{}0.482 \textpm{} 0.116} & {\scriptsize{}\cite{p92}}\tabularnewline
{\scriptsize{}28} & {\scriptsize{}SDSS-CMASS} & {\scriptsize{}0.59} & {\scriptsize{}0.488 \textpm{} 0.060} & {\scriptsize{}\cite{p94}}\tabularnewline
{\scriptsize{}29} & {\scriptsize{}BOSS DR12} & {\scriptsize{}0.38} & {\scriptsize{}0.497 \textpm{} 0.045} & {\scriptsize{}\cite{p2}}\tabularnewline
{\scriptsize{}30} & {\scriptsize{}BOSS DR12} & {\scriptsize{}0.51} & {\scriptsize{}0.458 \textpm{} 0.038 } & {\scriptsize{}\cite{p2}}\tabularnewline
{\scriptsize{}31} & {\scriptsize{}BOSS DR12} & {\scriptsize{}0.61} & {\scriptsize{}0.436 \textpm{} 0.034} & {\scriptsize{}\cite{p2}}\tabularnewline
{\scriptsize{}32} & {\scriptsize{}BOSS DR12} & {\scriptsize{}0.38} & {\scriptsize{}0.477 \textpm{} 0.051} & {\scriptsize{}\cite{p95}}\tabularnewline
{\scriptsize{}33} & {\scriptsize{}BOSS DR12} & {\scriptsize{}0.51} & {\scriptsize{}0.453 \textpm{} 0.050} & {\scriptsize{}\cite{p95}}\tabularnewline
{\scriptsize{}34} & {\scriptsize{}BOSS DR12} & {\scriptsize{}0.61} & {\scriptsize{}0.410 \textpm{} 0.044} & {\scriptsize{}\cite{p95}}\tabularnewline
{\scriptsize{}35} & {\scriptsize{}Vipers v7} & {\scriptsize{}0.76} & {\scriptsize{}0.440 \textpm{} 0.040} & {\scriptsize{}\cite{p55}}\tabularnewline
{\scriptsize{}36} & {\scriptsize{}Vipers v7} & {\scriptsize{}1.05} & {\scriptsize{}0.280 \textpm{} 0.080} & {\scriptsize{}\cite{p55}}\tabularnewline
{\scriptsize{}37} & {\scriptsize{}BOSS LOWZ } & {\scriptsize{}0.32} & {\scriptsize{}0.427 \textpm{} 0.056} & {\scriptsize{}\cite{p96}}\tabularnewline
{\scriptsize{}38} & {\scriptsize{}BOSS CMASS} & {\scriptsize{}0.57} & {\scriptsize{}0.426 \textpm{} 0.029} & {\scriptsize{}\cite{p96}}\tabularnewline
{\scriptsize{}39} & {\scriptsize{}Vipers} & {\scriptsize{}0.727} & {\scriptsize{}0.296 \textpm{} 0.0765} & {\scriptsize{}\cite{p97}}\tabularnewline
{\scriptsize{}40} & {\scriptsize{}6dFGS+SnIa} & {\scriptsize{}0.02} & {\scriptsize{}0.428 \textpm{} 0.0465} & {\scriptsize{}\cite{p98}}\tabularnewline
{\scriptsize{}41} & {\scriptsize{}Vipers} & {\scriptsize{}0.6} & {\scriptsize{}0.48 \textpm{} 0.12} & {\scriptsize{}\cite{p99}}\tabularnewline
{\scriptsize{}42} & {\scriptsize{}Vipers} & {\scriptsize{}0.86} & {\scriptsize{}0.48 \textpm{} 0.10 } & {\scriptsize{}\cite{p99}}\tabularnewline
{\scriptsize{}43} & {\scriptsize{}Vipers PDR-2} & {\scriptsize{}0.60} & {\scriptsize{}0.550 \textpm{} 0.120} & {\scriptsize{}\cite{p100}}\tabularnewline
{\scriptsize{}44} & {\scriptsize{}Vipers PDR-2} & {\scriptsize{}0.86} & {\scriptsize{}0.400 \textpm{} 0.110} & {\scriptsize{}\cite{p100}}\tabularnewline
{\scriptsize{}45} & {\scriptsize{}SDSS DR13} & {\scriptsize{}0.1} & {\scriptsize{}0.48 \textpm{} 0.16} & {\scriptsize{}\cite{p101}}\tabularnewline
{\scriptsize{}46} & {\scriptsize{}2MTF} & {\scriptsize{}0.001} & {\scriptsize{}0.505 \textpm{} 0.085} & {\scriptsize{}\cite{p102}}\tabularnewline
{\scriptsize{}47} & {\scriptsize{}Vipers PDR-2} & {\scriptsize{}0.85} & {\scriptsize{}0.45 \textpm{} 0.11} & {\scriptsize{}\cite{p103}}\tabularnewline
{\scriptsize{}48} & {\scriptsize{}BOSS DR12} & {\scriptsize{}0.31} & {\scriptsize{}0.469 \textpm{} 0.098} & {\scriptsize{}\cite{p49}}\tabularnewline
{\scriptsize{}49} & {\scriptsize{}BOSS DR12} & {\scriptsize{}0.36} & {\scriptsize{}0.474 \textpm{} 0.097} & {\scriptsize{}\cite{p49}}\tabularnewline
{\scriptsize{}50} & {\scriptsize{}BOSS DR12} & {\scriptsize{}0.40} & {\scriptsize{}0.473 \textpm{} 0.086} & {\scriptsize{}\cite{p49}}\tabularnewline
{\scriptsize{}51} & {\scriptsize{}BOSS DR12} & {\scriptsize{}0.44} & {\scriptsize{}0.481 \textpm{} 0.076} & {\scriptsize{}\cite{p49}}\tabularnewline
{\scriptsize{}52} & {\scriptsize{}BOSS DR12} & {\scriptsize{}0.48} & {\scriptsize{}0.482 \textpm{} 0.067} & {\scriptsize{}\cite{p49}}\tabularnewline
{\scriptsize{}53} & {\scriptsize{}BOSS DR12} & {\scriptsize{}0.52} & {\scriptsize{}0.488 \textpm{} 0.065} & {\scriptsize{}\cite{p49}}\tabularnewline
{\scriptsize{}54} & {\scriptsize{}BOSS DR12} & {\scriptsize{}0.56} & {\scriptsize{}0.482 \textpm{} 0.067} & {\scriptsize{}\cite{p49}}\tabularnewline
{\scriptsize{}55} & {\scriptsize{}BOSS DR12} & {\scriptsize{}0.59} & {\scriptsize{}0.481 \textpm{} 0.066} & {\scriptsize{}\cite{p49}}\tabularnewline
{\scriptsize{}56} & {\scriptsize{}BOSS DR12} & {\scriptsize{}0.64} & {\scriptsize{}0.486 \textpm{} 0.070} & {\scriptsize{}\cite{p49}}\tabularnewline
{\scriptsize{}57} & {\scriptsize{}SDSS DR7} & {\scriptsize{}0.1} & {\scriptsize{}0.376 \textpm{} 0.038} & {\scriptsize{}\cite{p104}}\tabularnewline
{\scriptsize{}58} & {\scriptsize{}SDSS-IV} & {\scriptsize{}1.52} & {\scriptsize{}0.420 \textpm{} 0.076} & {\scriptsize{}\cite{p105}}\tabularnewline
{\scriptsize{}59} & {\scriptsize{}SDSS-IV} & {\scriptsize{}1.52} & {\scriptsize{}0.396 \textpm{} 0.079} & {\scriptsize{}\cite{p106}}\tabularnewline
{\scriptsize{}60} & {\scriptsize{}SDSS-IV} & {\scriptsize{}0.978} & {\scriptsize{}0.379 \textpm{} 0.176} & {\scriptsize{}\cite{p107}}\tabularnewline
{\scriptsize{}61} & {\scriptsize{}SDSS-IV} & {\scriptsize{}1.23} & {\scriptsize{}0.385 \textpm{} 0.099} & {\scriptsize{}\cite{p107}}\tabularnewline
{\scriptsize{}62} & {\scriptsize{}SDSS-IV} & {\scriptsize{}1.526} & {\scriptsize{}0.342 \textpm{} 0.070} & {\scriptsize{}\cite{p107}}\tabularnewline
{\scriptsize{}63} & {\scriptsize{}SDSS-IV} & {\scriptsize{}1.944} & {\scriptsize{}0.364 \textpm{} 0.106} & {\scriptsize{}\cite{p107}}\tabularnewline
\hline
\end{tabular}
\end{table}

\begin{table}[ht]
\centering{}\caption{\label{t4}Hubbe parameter data \cite{R.Jimenez}.}

\renewcommand{\arraystretch}{0.45}

{\scriptsize{}}%
\begin{tabular}{ccc>{\centering}p{2cm}}
\hline
{\scriptsize{}Index} & {\scriptsize{}$z$} & {\scriptsize{}$H$} & {\scriptsize{}References}\tabularnewline
\hline
{\scriptsize{}1} & {\scriptsize{}0.0708} & {\scriptsize{}$69.0\pm19.68$} & {\scriptsize{}\cite{h69}}\tabularnewline
{\scriptsize{}2} & {\scriptsize{}0.09} & {\scriptsize{}$69.0\pm12.0$} & {\scriptsize{}\cite{h60}}\tabularnewline
{\scriptsize{}3} & {\scriptsize{}0.12} & {\scriptsize{}$68.6\pm26.2$} & {\scriptsize{}\cite{h69}}\tabularnewline
{\scriptsize{}4} & {\scriptsize{}0.17} & {\scriptsize{}$83.0\pm8.0$} & {\scriptsize{}\cite{h70}}\tabularnewline
{\scriptsize{}5} & {\scriptsize{}0.179} & {\scriptsize{}$75.0\pm4.0$} & {\scriptsize{}\cite{h71}}\tabularnewline
{\scriptsize{}6} & {\scriptsize{}0.199} & {\scriptsize{}$75.0\pm5.0$} & {\scriptsize{}\cite{h71}}\tabularnewline
{\scriptsize{}7} & {\scriptsize{}0.20} & {\scriptsize{}$72.9\pm29.6$} & {\scriptsize{}\cite{h69}}\tabularnewline
{\scriptsize{}8} & {\scriptsize{}0.27} & {\scriptsize{}$77.0\pm14.0$} & {\scriptsize{}\cite{h70}}\tabularnewline
{\scriptsize{}9} & {\scriptsize{}0.28} & {\scriptsize{}$88.8\pm36.6$} & {\scriptsize{}\cite{h69}}\tabularnewline
{\scriptsize{}10} & {\scriptsize{}0.35} & {\scriptsize{}$82.0\pm4.85$} & {\scriptsize{}\cite{h72}}\tabularnewline
{\scriptsize{}11} & {\scriptsize{}0.352} & {\scriptsize{}$83.0\pm14.0$} & {\scriptsize{}\cite{h73}}\tabularnewline
{\scriptsize{}12} & {\scriptsize{}0.3802} & {\scriptsize{}$83.0\pm13.5$} & {\scriptsize{}\cite{h73}}\tabularnewline
{\scriptsize{}13} & {\scriptsize{}0.4} & {\scriptsize{}$95.0\pm17.0$} & {\scriptsize{}\cite{h70}}\tabularnewline
{\scriptsize{}14} & {\scriptsize{}0.4004} & {\scriptsize{}$77.0\pm10.2$} & {\scriptsize{}\cite{h73}}\tabularnewline
{\scriptsize{}15} & {\scriptsize{}0.4247} & {\scriptsize{}$87.1\pm11.2$} & {\scriptsize{}\cite{h73}}\tabularnewline
{\scriptsize{}16} & {\scriptsize{}0.4497} & {\scriptsize{}$92.8\pm12.9$} & {\scriptsize{}\cite{h73}}\tabularnewline
{\scriptsize{}17} & {\scriptsize{}0.4783} & {\scriptsize{}$80.9\pm9.0$} & {\scriptsize{}\cite{h73}}\tabularnewline
{\scriptsize{}18} & {\scriptsize{}0.48} & {\scriptsize{}$97.0\pm62.0$} & {\scriptsize{}\cite{h74}}\tabularnewline
{\scriptsize{}19} & {\scriptsize{}0.593} & {\scriptsize{}$104.0\pm13.0$} & {\scriptsize{}\cite{h71}}\tabularnewline
{\scriptsize{}20} & {\scriptsize{}0.68} & {\scriptsize{}$92.0\pm8.0$} & {\scriptsize{}\cite{h71}}\tabularnewline
{\scriptsize{}21} & {\scriptsize{}0.781} & {\scriptsize{}$105.0\pm12.0$} & {\scriptsize{}\cite{h71}}\tabularnewline
{\scriptsize{}22} & {\scriptsize{}0.875} & {\scriptsize{}$125.0\pm17.0$} & {\scriptsize{}\cite{h71}}\tabularnewline
{\scriptsize{}23} & {\scriptsize{}0.88} & {\scriptsize{}$90.0\pm40.0$} & {\scriptsize{}\cite{h74}}\tabularnewline
{\scriptsize{}24} & {\scriptsize{}0.9} & {\scriptsize{}$117.0\pm23.0$} & {\scriptsize{}\cite{h70}}\tabularnewline
{\scriptsize{}25} & {\scriptsize{}1.037} & {\scriptsize{}$154.0\pm12.0$} & {\scriptsize{}\cite{h71}}\tabularnewline
{\scriptsize{}26} & {\scriptsize{}1.3} & {\scriptsize{}$168.0\pm17.0$} & {\scriptsize{}\cite{h70}}\tabularnewline
{\scriptsize{}27} & {\scriptsize{}1.363} & {\scriptsize{}$160.0\pm33.6$} & {\scriptsize{}\cite{h75}}\tabularnewline
{\scriptsize{}28} & {\scriptsize{}1.43} & {\scriptsize{}$177.0\pm18.0$} & {\scriptsize{}\cite{h70}}\tabularnewline
{\scriptsize{}29} & {\scriptsize{}1.53} & {\scriptsize{}$140.0\pm14.0$} & {\scriptsize{}\cite{h70}}\tabularnewline
{\scriptsize{}30} & {\scriptsize{}1.75} & {\scriptsize{}$202.0\pm40.0$} & {\scriptsize{}\cite{h70}}\tabularnewline
{\scriptsize{}31} & {\scriptsize{}1.965} & {\scriptsize{}$186.5\pm50.4$} & {\scriptsize{}\cite{h75}}\tabularnewline
\hline
\end{tabular}{\scriptsize\par}

\end{table}

\newpage{}

\end{document}